\documentclass[prd,superscriptaddress,nofootinbib,colorlinks=true,preprint]{revtex4-2}
\usepackage{amsmath, amssymb, graphicx, hyperref, slashed}
\usepackage[dvipsnames]{xcolor}

\pdfoutput=1 

\def\lag{\mathcal{L}}

\newcommand\Tstrut{\rule{0pt}{3.5ex}}         
\newcommand\Bstrut{\rule[-2ex]{0pt}{0pt}}   

\def\beq{\begin{equation}}
\def\eeq{\end{equation}}

\def\bit{\begin{itemize}}
\def\eit{\end{itemize}}
\def\ben{\begin{enumerate}}
\def\een{\end{enumerate}}

\def\lag{\mathcal{L}}

\def\Msolar{M_\odot}
\def\eV{\,\text{eV}}

\def\GeV{\,\text{GeV}}

\def\cm{\,\text{cm}}
\def\km{\,\text{km}}
\def\pc{\,\text{pc}}
\def\kpc{\,\text{kpc}}
\def\m{\,\text{m}}
\def\s{\, \text{s}}
\def\yr{\,\text{yr}}
\def\Gyr{\, \text{Gyr}}

\def\K{\, \text{K}}

\def\rhoDM{\rho_{X}}

\def\mn{m_n}

\def\af{\alpha_n}

\def\alf{\tilde \alpha}

\def\red{\text{red}}

\def\Eq#1{Eq.~\ref{#1}}
\def\Fig#1{Fig.~\ref{#1}}
\def\Tab#1{Table~\ref{#1}}
\def\Sec#1{Sec.~\ref{#1}}
\def\App#1{Appendix~\ref{#1}}

\begin{document}

\title{Astrophysical Observations of a Dark Matter-Baryon Fifth Force}

\author{Moira I. Gresham}
\affiliation{Physics Department, Whitman College,
	Boyer Ave, Walla Walla, WA, U.S.A.}
\author{Vincent S. H. Lee}
\affiliation{Walter Burke Institute for Theoretical Physics, California Institute of Technology,
	California Blvd, Pasadena, CA, U.S.A.
}
\author{Kathryn M. Zurek}
\affiliation{Walter Burke Institute for Theoretical Physics, California Institute of Technology,
	California Blvd, Pasadena, CA, U.S.A.
}

\preprint{CALT-TH-2022-032}

\begin{abstract} 
\noindent We consider the  effects of an attractive, long-range Yukawa interaction between baryons and dark matter (DM), focusing in particular on temperature and pulsar timing observations of neutron stars (NSs). We show that such a fifth force, with strength modestly stronger than gravity at ranges greater than tens of kilometers (corresponding to mediator masses less than $10^{-11} \text{eV}$), can dramatically enhance dark matter kinetic heating, capture, and pulsar timing Doppler shifts relative to gravity plus short range interactions alone. Using the coldest observed NS and pulsar timing array (PTA) data, we derive limits on fifth force strength over a DM mass range spanning light dark matter up to order solar mass composite DM objects. We also consider an indirect limit by combining bullet cluster limits on the DM self-interaction with weak equivalence principle test limits on baryonic self-interactions. We find the combined indirect limits are moderately stronger than kinetic heating and PTA limits, except when considering a DM subcomponent. 
\end{abstract}

\maketitle
\newpage
\tableofcontents
\newpage


\section{Introduction}\label{sec: introduction}
A long-range fifth force between baryonic matter has long been the focus of both theoretical and experimental inquiry, as reviewed in Ref.~\cite{Adelberger_2003}.  If DM also interacts with baryons via an attractive fifth force, it induces a potential, 
\beq
V_\text{Yuk}(r) = -\tilde{\alpha}{G M m_X \over r }  e^{- r / \lambda}, \label{eq: yukawa}
\eeq
where $m_X$ and $M$ are the masses of the DM and a macroscopic baryonic object, respectively, when the sizes of both objects are much smaller than the force range, $\lambda$, and separation, $r$. 
This potential can arise from an effective interaction $\lag \supset g_X \phi \bar XX + g_n \phi \bar n n$ where $X$ and $n$ are the effective DM and nucleon fields, and $\phi$ can be either a massive but very light scalar or vector field.  The effective coupling in this simplified model is 
\beq \tilde \alpha \approx {g_n g_X \over 4\pi G m_n m_X},\eeq  
where the approximation holds well when the fifth force and gravitational binding energies are subdominant contributors to the mass, $M$.  Here we have in mind that $m_X$ could be the mass of a DM particle, or a macroscopic DM object such as a nugget of asymmetric DM~\cite{Hardy:2014mqa,Krnjaic:2014xza,Detmold:2014qqa,Wise:2014jva,Gresham:2017cvl,Gresham:2017zqi,Gresham:2018anj,Gresham:2018rqo}. 

\begin{figure}
\includegraphics[width=\textwidth]{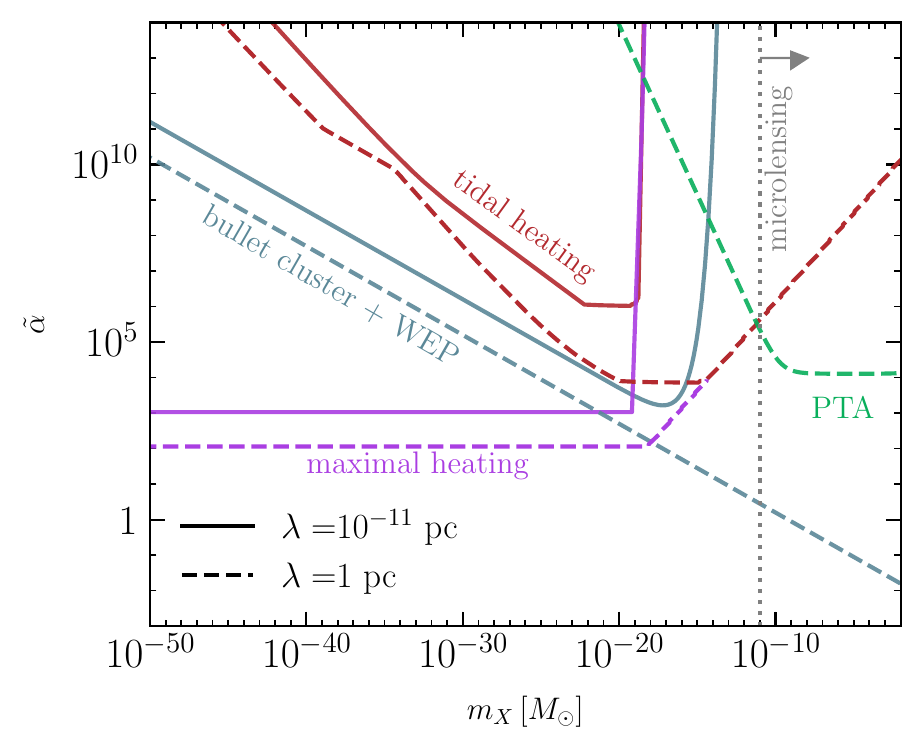}
\caption{Summary of the bounds derived in this paper on the DM-baryon fifth force to gravitational force strength ratio, $\alf$, as defined in \Eq{eq: yukawa}, for force ranges, $\lambda$, at the two extremes of the range we consider. The relevant equations for tidal heating, maximal heating, PTA, and bullet cluster $+$ WEP are derived in Sec.~\ref{subsec: graze}-\ref{subsec: captured}, Sec.~\ref{subsec: max capture}, Sec.~\ref{sec: PTA}, and Sec.~\ref{subsec:sidm} respectively. Bound from microlensing surveys is shown in a gray line, which apply to compact DM with radii less than $\sim0.1$ solar radius~\cite{Croon:2020ouk}.}\label{fig: money}
\end{figure}

The goal of this paper is to consider the astrophysical observables of such a DM-baryon fifth force~\footnote{A recent work on long-range DM-baryon fifth force with a different set-up can be found in~\cite{Kim_2021}.}, focusing on a few simple tests that constrain such an interaction.  Focusing on astrophysical tests implies that we are interested in force ranges $20 \km \ll \lambda \ll \kpc$, corresponding to (ultralight) mediator masses $10^{-11} \eV \gg m_\phi \gg 10^{-26} \eV$.\footnote{Above the $\kpc$ scale, torsion balance tests of differential accelerations toward the galactic center limit the baryon-DM force to be weaker than gravity. The lower limit is set by neutron star diameters.}  Firstly, one must consider the constraints separately on $g_n$ and $g_X$, which are dominated by the MICROSOPE mission's weak equivalence principle (WEP) test~\cite{Berge:2017ovy,Fayet:2018cjy} and DM self-interactions \cite{Spergel:1999mh,Kahlhoefer:2013dca,Coskuner:2018are}, respectively. Combining these constraints allows one to derive a bound on $\tilde \alpha$, shown in Fig.~\ref{fig: money} as ``bullet cluster + WEP'' for two different astrophysical force ranges.  Note, importantly, that this combination of bounds will lift quickly for a DM sub-component, only weakly constrained by observations of DM halos.

The majority of this paper will focus on a pair of constraints that weaken only linearly with the DM density for a DM sub-component.  These come from heating of neutron stars (NSs) from DM capture, and pulsar timing measurements of transiting DM clumps, where DM passing near pulsars enhance the Doppler effect on the pulsar frequencies \cite{Siegel_2007, Seto_2007,  Kashiyama_2012, Clark_2015_I, Clark_2015_II, Schutz_2017, Kashiyama_2018, Baghram:2011is, Dror:2019twh, Ramani:2020hdo, Lee_2021, lee2021bayesian}. These two constraints on the DM-baryon fifth force, which we derive in detail below, are summarized in Fig.~\ref{fig: money}, labeled as ``heating'' and ``PTA.''   The heating constraints are further shown for two different limits: first, where the DM interactions beyond gravity are only via the fifth force (labeled as ``tidal''), and, second, where the DM has not only the fifth force to focus it onto the NS but also a short range interaction to capture it with high efficiency (labeled as ``maximal''). In Fig.~\ref{fig: money} we also show constraints from microlensing surveys, which rule out DM with $M>10^{-11}\,M_{\odot}$ and radii less than $\sim0.1$ solar radius~\cite{Croon:2020ouk}. This is relevant for our kinetic heating analysis, since we assume that each DM is smaller than the size of a typical NS. The PTA constraints, however, are unaffected by the microlensing bounds, since the PTA analysis only assumes DM to be smaller than the impact parameter relative to the NS, which are at least $b>2.5\times10^4$ solar radii for $M>10^{-11}\,M_{\odot}$ (cf. Eq.~\eqref{eq:bmin}), and thus cannot be effectively constrained by microlensing studies.

Heating of NSs via DM capture has been considered previously for the case of a short-range interaction, such as for WIMPs and hidden sector DM \cite{Baryakhtar:2017dbj,Raj:2017wrv, Garani:2018kkd, Bell:2018pkk, Bell:2019pyc, Garani:2019fpa, Acevedo:2019agu, Joglekar:2019vzy, Dasgupta:2020dik, Keung:2020teb, Joglekar:2020liw, Bell:2020obw, Bell:2020lmm, Bell:2020obw,Garani:2020wge, Maity:2021fxw,McKeen:2021jbh,Ilie:2021iyh, Ilie:2021umw, Bell:2020jou, Bramante:2021dyx}. However, when the force range is longer than a typical NS diameter, $\lambda > 20\,\mathrm{km}$  ($m_\phi < 10^{-11} \eV $), 
a DM-baryon force accelerates and focusses DM more than through gravity alone, leading to observable effects via two different mechanisms which we compute in detail. First, more DM is focussed on the NS due to the fifth force.  Second, the DM is more energetic when it arrives at the NS surface, which we show then dominantly heats the NS via seismic oscillations.   We then extract the heating constraints utilizing the coldest observed NS (PSR J2144-3933, from Hubble Space Telescope) by requiring that the kinetic energy of all the captured DM raise the temperature by an amount less than the observed (unredshifted surface) temperature, $T_s < 42,000$ K \cite{Guillot_2019}.  

The rest of the paper is devoted to deriving in detail the constraints on the DM-baryon fifth force coupling $\tilde \alpha$ outlined in Fig.~\ref{fig: money}. In Sec.~\ref{sec: kinetic heating}, we derive the kinetic heating rate and resulting NS luminosity in the presence of a fifth force interaction. Some of the details  are relegated to  Appendices.  In Sec.~\ref{sec: PTA}, we outline the procedure for deriving fifth force constraints with PTA observations, and show the results using the  11-year dataset~\cite{Arzoumanian_2018_data, Arzoumanian_2018_GW} by NANOGrav~\cite{brazier2019nanograv}.   In Sec.~\ref{sec: wep bc}, we consider observations from WEP tests and the bullet cluster, and derive  indirect upper limits on the DM-baryon interaction. Finally, in Sec.~\ref{sec: conclusion}, we conclude.

\section{Limits from NS Temperature Observations}\label{sec: kinetic heating}

Transiting DM can heat NSs to higher-than-expected temperatures observable by the next generation of infrared telescopes, including the James Webb Space Telescope (JWST) \cite{Baryakhtar:2017dbj}. As discussed in \cite{Baryakhtar:2017dbj}, if the energy deposited by DM passing near a NS is dissipated through black-body radiation, then in the local rest frame of the NS, the thermal photon luminosity due to dark kinetic heating is $L_\gamma^\text{DM} = \dot{E}_\text{kin}$. Given an unredshifted surface temperature measurement $T_{\text{s}, i} \leq {T_{\text{s}, i}}_\text{max}^{\text{meas}}$ for a particular NS, $i$, with radius, $R$, we can infer a limit on the factors contributing to $\dot E_\text{kin}$ through\footnote{Redshifted surface temperature, ${T^\infty}$, is related to unredshifted temperature, $T$, through $T = {T^\infty \over \sqrt{1 - {2 G M \over R}}}$.}
\beq
\dot E_\text{kin} <  {\sigma_\text{B} 4 \pi R^2   \left({T_{\text{s}, i}}_\text{max}^{\text{meas}}\right)^4 }, \label{eq: limit}
\eeq
where the heating rate is schematically given by
\beq
\dot E_\text{kin} \approx \dot N \Delta E, \label{eq: heating_rate_schematic}
\eeq
with $\dot N$ and $\Delta E$ being the relevant DM flux and single-DM energy transfer, respectively. 

All previous analyses of DM kinetic heating of NSs have focused on short-range DM interactions. As explained in \cite{Baryakhtar:2017dbj}, in this case, DM can deposit energy on a NS only if it intersects the NS, requiring impact parameters less than $b_{\max}^{\mathrm{grav}}=v^{-1}\sqrt{2GM R/(1-2GM/R)}$. At a maximum, an order one fraction of DM that intersects the NS is captured and deposits all of its kinetic energy at the NS surface, also known as the gravitational binding energy released in the capture process. Hence $\Delta E^{\mathrm{grav}}_\text{max}\approx GMm_X/R$ and $\dot N^{\mathrm{grav}}_\text{max}\approx(\rhoDM/m_X)\pi \langle b_{\max}^2 v\rangle$, leading to a maximum heating rate from DM focused by gravity onto typical NSs near our galactic radius (${G M / R} \sim 0.2$, $\langle v \rangle \sim 10^{-3}$, $\rhoDM \sim 0.4 \GeV/\cm^3$),
\beq 
\dot E^{\mathrm{grav}}_\text{kin, max} \approx {G M m_X \over R}{\pi {\rhoDM \over m_X}} \langle \left({b_{\max}^{\mathrm{grav}}}\right)^2 v \rangle \approx 4 \pi R^2 \sigma_\text{B} (2000 \text{K})^4. \label{eq: max heating rate}
\eeq
Therefore measured (unredshifted) surface temperatures of nearby NSs below about $2000 \text{K}$ can start to constrain short-range interactions of DM with NS matter. Conversely, finding several old isolated neutron stars with temperatures of order $2000 \text{K}$ and greater near our Galactic radius---with higher temperatures in DM-rich regions---could be a sign of DM kinetic heating.

We now consider a long-range ($> 20 \km$) fifth force, which changes the heating rate calculation, Eq.~\eqref{eq: heating_rate_schematic}, in three ways. First, more DM is focussed and potentially captured by the NS (larger $b_\text{max}$).\footnote{A repulsive interaction would defocus DM and reduce the maximum possible kinetic heating rate. Here we focus on the attractive case.} Second, the DM's kinetic energy near the NS surface---roughly equivalent to the DM-NS binding energy---is larger, meaning more energy is dissipated in a DM capture process (larger $\Delta E$).\footnote{Similar effects on Earth could be relevant for direct detection, and were discussed in \cite{Davoudiasl:2017pwe, Davoudiasl:2020ypv}.} Third, the energy deposition mechanism is qualitatively different due both to the long-range force and ultrarelativistic DM speeds at the NS surface; collective excitations of NS matter are particularly important.

We divide DM  contributing to heating into two groups:
\begin{itemize}
	\item[A.] \emph{Grazing} DM, including NS-transiting DM that is not gravitationally captured, as well as gravitationally bound DM whose first orbital period after  capture is greater than the NS lifetime, implying no subsequent energy deposits within the NS's lifetime.
	\item[B.] \emph{Captured} DM, which deposits sufficient energy to become gravitationally bound on its first transit and deposits further energy on subsequent transits within the NS lifespan.
\end{itemize}
The total heating rate for a long-range interaction is thus
\beq
\dot E_\text{kin} \approx  \dot E_\text{graze} + \dot E_\text{cap} \label{eq: Ekin tidal}
\eeq
with $\dot E_\text{graze} \approx \dot N_{\text{graze}}\Delta E_{\text{graze}}$ and $\dot E_\text{cap} \approx \dot N_{\text{cap}}\Delta E_{\text{cap}}$. Below in Secs.~\ref{subsec: graze}-\ref{subsec: captured}, we derive explicit estimates of $\Delta E_{\mathrm{graze}}$, $\dot N_{\mathrm{graze}}$, $\Delta E_{\mathrm{cap}}$, and $\dot N_{\mathrm{cap}}$ in Eqs.~\ref{eq: dE}, \ref{eq: N dot graze}, \ref{eq: cap dE}, and \ref{eq: cap N dot}, respectively. Then in \Sec{subsec: max capture} we discuss the maximal heating rate, analogous to \Eq{eq: max heating rate}, requiring assistance by additional short-range forces. 
In \Sec{subsec: continuous heating} we estimate when heating is effectively continuous and possibly destructive to the NS, which is relevant at the higher mass scales we consider. Finally, in \Sec{subsec: limits}, we present limits based on our estimates and \Eq{eq: limit}.


\subsection{Grazing heating rate}\label{subsec: graze}
\subsubsection{$\Delta E_{\mathrm{graze}}$} \label{subsec: dE graze}
Two stars in a close encounter can become bound by sinking energy and angular momentum into seismic oscillations of the stars through gravitational tidal forces \cite{1977ApJ...213..183P, 1975MNRAS.172P..15F}. Pani and Loeb \cite{2014} considered a similar capture mechanism for primordial black holes (PBHs) by NSs. They modeled  PBHs as point-like objects that remain intact as they transit the NS, and found that tidally deposited energy exceeds energy deposit through dynamical friction by several orders of magnitude. A small, compact DM object should behave similarly, and an additional Yukawa force with range much larger than the NS radius will simply effectively increase the strength of the gravitational tidal forces by a factor $1 + \tilde{\alpha}$. 

Based on \cite{2014}, we estimate the energy deposited through tidal excitation of NS seismic oscillations by DM that transits an NS as
\beq
\Delta E_{\mathrm{graze}} \approx {G m_X^2 (1 + \tilde{\alpha})^2 \over R} 4 \sqrt{l_\text{max}} \label{eq: dE}
\eeq where $l_\text{max}$ corresponds to the largest spherical seismic mode excited in the NS. 
Qualitatively, thinking of the NS like a spring, the energy goes as the square of the amplitude of the oscillation, which is set by the maximum tidal force at close approach---proportional to $(1 + \tilde{\alpha}) m_X$.   The stiffness of the NS equation of state determines the nontrivial dependence on mode number. We refer the reader to \cite{2014} for details.

Ref.~\cite{2014} considers many possible cutoffs for $l_\text{max}$. The limiting cutoffs come from demanding the particle size, $R_X$, is smaller than the mode's wavelength, leading to $l_\text{max} < R / R_{X}$, and that the DM crossing time, $\tau_\text{cross} \sim R/\beta_R \sim R$, is short compared to the shear viscosity oscillation damping timescale \cite{1987ApJ...314..234C},
\beq
\tau_\eta \approx {\rho R^2 / \eta \over (l-1)(2 l+1)}, \label{diss 1} \\
\eeq
where $\eta$ is the shear viscosity, and $\rho$ is the mass density of the NS. For cold NSs with $T < 10^8$K, in which  neutrons and protons are expected to be superfluid (at least outside the inner core of the NS, where the tidally-induced oscillations are supported) the dominant source of viscosity is from electron scattering, $\eta_e$ \cite{Andersson:2006va}. To within an order of magnitude, for a given temperature, the maximal value of $\eta_e / \rho$ occurs around $\rho \sim 10^{10}$--$4\times10^{14} {\text{g} \over \text{cm}^3}$, and is \cite{1976ApJ...206..218F,1979ApJ...230..847F}
\beq
\left(\eta_e \over \rho \right)_\text{max} \approx \left(T \over 10^8 \text{K} \right)^{-2}10^4 \text{cm}^2/\text{sec},
\eeq     
leading to an estimate
\beq
l_\text{max} \sim \min \left( 10^2 \left(T \over 10^4 \text{K} \right), {R \over R_X} \right) \label{eq: lmax}
\eeq for DM that is relativistic (and ultrarelativistic) at the NS surface, 
with the consistency requirement, $l_\text{max} > 1$.
For simplicity and consistency, we consider only DM and NSs with 
\beq {R_X \over R} < 10^{-2} {10^4 \text{K} \over T} <  1 \qquad \text{and} \qquad T<10^8 \text{K} \label{eq: consistency}\eeq such that the shear viscosity timescale determines the highest excited mode number. For maximally compact (i.e. $\frac{Gm_X}{R}\sim 1/2$) DM objects, this requires $m_X \lesssim 10^{-2} {10^4 \text{K} \over T} M_\odot < M_\odot $. The cutoff occurs at lower mass for less compact DM.

A few other factors deserve mention. First, our estimate based on \cite{2014} is consistent only if  both the DM and NS survive the DM's transit. For DM to survive, the Roche distance---where the tidal forces on the DM from the NS start to exceed the binding forces---must be smaller than the NS radius. Depending on the forces binding the DM, the DM must be sufficiently compact.\footnote{Modeling the DM binding force per mass as proportional to ${({\tilde\alpha}_{\chi,\text{eff}} + 1) G m_X \over R_X^2}$ on the DM surface, given $R_X \ll R$, the Roche distance is approximately $r_\text{Roche} \sim \left(4 {(1 + \tilde{\alpha}) M \over (1 + \tilde{\alpha}_{\chi,\text{eff}}) m_X}\right)^{1/3} R_X$, and in terms of the DM and NS compactnesses, the DM can survive only if ${G m_X \over R_X}> 4 {(1 + \tilde{\alpha}) \over (1 + \tilde{\alpha}_{\chi,\text{eff}})}{G M \over R} \left({R_X \over R}\right)^2$. } 
Energy deposits comparable to a few percent of the NS's gravitational binding energy could also destroy the NS. We discuss this case further in Secs.~\ref{subsubsec: delta E cap} and \ref{subsec: continuous heating}.

Second, the estimate \Eq{eq: dE} is based on analyzing a particle's infall from rest. But as discussed in \cite{2014}, it should be a decent estimate for particles with other trajectories as long as they breach or come close to breaching the NS surface. On the other hand, at distances, $r$, large compared to $R$, the tidal force falls off as $r^{-3}$ and effective crossing times lengthen when $r_\text{min} \gg R$. Indeed estimates of gravitational tidal capture for close star encounters as in \cite{1977ApJ...213..183P} apply in this case, and the moral is that a good estimate of heating and capture comes from counting only DM that intersects the NS, with $\Delta E_{\mathrm{graze}}$ as in Eqs.~\ref{eq: dE} and \ref{eq: lmax} for all such DM.

\subsubsection{$\dot N_{\mathrm{graze}}$}
The number of DM particles of mass $m_X$ and asymptotic mass density $\rhoDM$ passing through a given NS per time---the flux---is
\beq \dot N_{\mathrm{graze}} = {\rhoDM \over m_X} \pi \langle b_\text{max}^2 v \rangle.
\label{eq: N dot graze}
\eeq
Given focusing through gravity alone, $\dot N_{\mathrm{graze}} \sim {10^{-24} \over \yr} {\rhoDM \over 0.4 \GeV/\cm^3}\left\langle {10^{-3} \over v }\right\rangle {M_\odot \over m_X} {R \over 10 \km}{G M \over 2 \km}$. 
An attractive DM-NS fifth force focuses more DM onto a NS, leading to a significantly larger flux via  larger $b_\text{max}$. For moderate $\tilde{\alpha}$ and $\lambda \gg R$, compared to focusing through gravity alone, $b_\text{max}$ is larger by a factor $\sqrt{1+ \alf}$. When $\tilde{\alpha} \gg 1$, $b_\text{max}$ grows linearly with $\tilde{\alpha}$ until a cutoff where an \emph{outer} centrifugal barrier at $r > \lambda$ from the exponential turn-on of the fifth force becomes stronger than the always-present \emph{inner} centrifugal barrier. More specifically, when $\alf \gg 1$,  
\beq
{b_\text{max} v}  \sim \Bigg\{ 
\begin{array}{l l} 
{G M} \alf & \text{inner barrier}\\
{\sqrt{2 G M \lambda  \log\left( \alf \log \alf \right)}} & \text{outer barrier when}~\alf e^{-G M / \lambda v^2} < 1 \\
{\lambda v} \log\left( {\alf G M  \over \lambda v^2} \right) & \text{outer barrier otherwise},\\
\end{array} \label{eq: bmax large alpha approx}
\eeq
with $b_{\max}=\min(b_{\max,\,\rm{inner}},b_{\max,\,\rm{outer}})$ if $\alf GM/\lambda v^2\gtrsim e$ and $b_{\max}=b_{\max,\,\rm{inner}}$ otherwise. In \App{app: orbits}, we derive the general relativistic equation of motion that determines $b_\text{max}$ in Eqs.~\ref{eq: orbit eom}-\ref{eq: bmax numerical}
along with analytic approximations in Eqs.~\ref{eq: b inner}-\ref{eq: b};  \Fig{fig: bmax} shows our analytic approximations alongside numerical solutions to the exact barrier-determining expressions.

\subsection{Captured heating rate}\label{subsec: captured}
\subsubsection{$\Delta E_{\mathrm{cap}}$} \label{subsubsec: delta E cap}

The energy deposited per DM particle captured is approximately the DM kinetic energy at the NS surface, \beq E_\text{DM, kin}(R) \approx m_X (\gamma_R - 1). \label{eq EDMkin} \eeq 
with
\beq \gamma_R = {{1 \over \sqrt{1 - v^2}} + \tilde{\alpha}{G M \over R} e^{-R/\lambda} \over \sqrt{1 - {2 G M \over R}}} \approx {1 + {G M \over R} \tilde{\alpha} e^{-R/\lambda} \over \sqrt{1 - {2 G M \over R}}} \approx 1 + {G M \over R}\left(1 + \tilde{\alpha} e^{-R/\lambda} \right) \label{eq gammaR} \eeq 
as measured in a locally flat frame at the NS surface. Since DM asymptotic speeds, $v$, are generally much less than the escape speed at the NS surface, the energy deposit is essentially independent of $v$.
Without a fifth force, $\gamma_R$ is given by Eq.~\eqref{eq gammaR} but with $\tilde{\alpha}=0$, which for typical neutron stars and DM with $v\ll 1$, is $\gamma_R \sim (1-\frac{2GM}{R})^{-1/2} \sim \sqrt{5/3}$. With a fifth force, $\gamma_R - 1$ is larger by about a factor of $1 + \tilde{\alpha} e^{-R/\lambda}$. 

If the period of the first orbit after capture, $\Delta t_1$, is greater than NS lifetime, then the DM deposits only energy $\Delta E_\text{graze}$ within the NS lifetime. Hence, we cannot simply use the expression for $E_\text{DM, kin}(R)$ in Eq.~\eqref{eq EDMkin} for $\Delta E_{\mathrm{cap}}$. Instead, we use an empirical formula for $\Delta E_{\mathrm{cap}}$ to smoothly interpolate between different limits of $\Delta t_1/t_{\mathrm{NS}}$, written as
\beq
\Delta E_{\text{cap}} \approx \left({\Delta E_{\mathrm{graze}} \over E_\text{DM, kin}(R) }\right)^{\Delta t_1 / t_\text{NS}} E_\text{DM, kin}(R)  \qquad (\text{when} \, \Delta E_\text{graze} \ll E_\text{DM, kin}(R)) \, .\label{eq: cap dE}
\eeq 
We now justify this expression. If $\Delta t_1\gg t_{\mathrm{NS}}$, then DM would not have enough time to deposit energy into NS, and we expect $\Delta E_{\text{cap}}\to 0$. In the opposite limit where $\Delta t_1\ll t_{\mathrm{NS}}$, DM has ample time to transfer all its energy and be completely captured by the NS, hence one expects $\Delta E_{\text{cap}}= E_\text{DM, kin}(R)$. Finally, if $\Delta t_1\sim t_{\mathrm{NS}}$, then we expect DM to deposit some but not all of its kinetic energy to the NS, since the DM is not fully captured by NS. Effectively, the DM grazes the NS once. It is thus appropriate to approximate $\Delta E_{\text{cap}}\sim\Delta E_{\text{graze}}$ in this scenario. One can easily check that the empirical formula in Eq.~\eqref{eq: cap dE} is a smooth function that reduces to these three limits. In the case where $\Delta t_1\gtrsim t_{\mathrm{NS}}$, the DM is counted as ``grazing'' DM.\footnote{Note the first energy deposit of captured DM is double counted in $\dot E_\text{graze}$ and $\dot E_\text{cap}$, which is acceptable for an order-of-magnitude estimate since it will at most overestimate the heating effect by a factor of two, and in most cases by a only a tiny fraction over one.}
We estimate the orbital period in \App{sec: timescale for heating}; see $\Delta t_1$ in Eqs.~\eqref{eq: orbit period}-\eqref{eq: rmax1}. 

Using Eqs.~\ref{eq: dE}, \ref{eq: lmax}, \ref{eq EDMkin}, and \ref{eq gammaR}, we see $\Delta E_\text{graze} < E_\text{DM, kin}(R)$ as long as ${m_X \over M}(1 + \alf) 40 \sqrt{T \over 10^4 \text{K}} < 1$. The condition saturates when $\Delta E_\text{graze} \approx E_\text{DM, kin}(R) \approx \sqrt{10^4 \text{K} \over T}{1 \over 40}{G M^2 \over R}$, or when the energy deposit is about 10\% of the NS binding energy ($\sim {3 \over 5}{G M^2 \over R}$) given $T \sim 42,000\text{K}$. We expect such an energy deposit to  destroy the NS rather than ``heat'' it, as we will discuss further in \Sec{subsec: continuous heating}. Therefore $\Delta E_\text{graze}< E_\text{DM, kin}(R)$  holds as long as the kinetic heating limit is relevant.

\subsubsection{$\dot N_{\mathrm{cap}}$} 
To become gravitationally bound to the NS, DM must lose sufficient energy when it first grazes the NS surface. Thus the capture rate is approximately 
\beq
\dot N_\text{cap} \approx {\rhoDM \over m_X} \pi \langle b_\text{max}^2 v \rangle_{v < v_\text{cap}}, 
\eeq where the average is over asymptotic DM speeds up to a the maximum speed of DM that is captured: 
\beq v_\text{cap} = \sqrt{2 \Delta E_{\mathrm{graze}} / m_X}. \label{eq: vmaxcap}\eeq 
Given a Maxwellian velocity distribution with peak speed $v_p$, since $(b_\text{max} v)$ is nearly constant as function of $v$---increasing negligibly for most or our parameter range of interest and less than linearly in the entire range---to reasonable approximation,
\beq
 \dot N_\text{cap} \approx {\rhoDM \over m_X} \pi (b_\text{max} v)^2|_{v\rightarrow \min(v_p, v_\text{cap})} {2 \over \sqrt{\pi}  v_p} f \label{eq: cap N dot}
\eeq
where
\beq
f \approx \left(1 - e^{ - {v^2_\text{cap}}/v_p^2}\right)  \label{eq: f app}
\eeq
roughly represents the fraction of surface-breaching DM that is captured. We can also use Eq.~\eqref{eq: cap N dot} to estimate $\dot N_{\mathrm{graze}}$ by taking $v_\text{cap}\to 1$.


\subsection{Capture assisted by short-range interactions}\label{subsec: max capture}

We have just described a capture mechanism through DM's tidal force excitation of seismic oscillations in an NS---classical collective modes stretching over the entire NS. In contrast, most previous treatments of kinetic heating have focussed on effectively local interactions of DM with one or more individual nucleons or leptons within the NS during a transit \cite{Baryakhtar:2017dbj,Raj:2017wrv, Garani:2018kkd, Bell:2018pkk, Bell:2019pyc, Garani:2019fpa, Acevedo:2019agu, Joglekar:2019vzy, Dasgupta:2020dik, Keung:2020teb, Joglekar:2020liw, Bell:2020obw, Bell:2020lmm, Bell:2020obw,Garani:2020wge, Maity:2021fxw,McKeen:2021jbh,Ilie:2021iyh, Ilie:2021umw, Bell:2020jou, Bramante:2021dyx}. A notable exception is \cite{derocco2022dark}, which focuses specifically on collective effects within dense stellar media in DM scattering.

Relative to the case without a fifth force, when the fifth force increases DM's speeds in the NS rest frame (c.f.~\Eq{eq gammaR}), the kinematic upper limit on energy transfer to nucleons in elastic DM-nucleon interactions also increases. At the same time, length contraction of the NS in the DM-nucleon CM frame can be significant, so it is easier for the de Broglie wavelength of the CM motion ($q^{-1} \sim 1/\sqrt{2 m_n \Delta E_{\mathrm{graze}}}$) to be larger than the inter-nucleon distance along the direction of motion in the CM frame. Thus while the increased CM energy might naively increase nucleon-DM elastic cross sections, the range of kinematically allowed energies where nucleons are effectively free is also narrower. When the DM is ultrarelativistic at the NS surface, we expect proper treatment of collective effects in NS matter to be particularly important in a treatment of DM scattering via short-range forces. 

Rather than trying to model specific short-range interactions and account for the relevant microscopic NS physics that determines the probability of capture, we consider maximal heating, where an order one fraction of transiting DM is captured. In this case, 
\beq
\dot E_\text{kin, max} \approx E_\text{DM, kin}(R) \dot N_\text{graze} \label{eq: Ekinmax}
\eeq
with $E_\text{DM, kin}(R)$ given by \Eq{eq EDMkin} and $\dot N_\text{graze}$ by \Eq{eq: N dot graze}.

\subsection{Cooling timescale and continuous heating approximation}\label{subsec: continuous heating}

NSs are born with internal temperatures of order $10^{11}~\text{K}$ and rapidly cool through neutrino and photon emission to under $10^7~\text{K}$ within about 1000 years. Without additional sources of heating, NSs are expected to cool to under 1000~\text{K} within about 20 Myr (see {\em e.g.} \cite{Page:2004fy, Yakovlev_2004}).   Under temperatures of order $10^6~\text{K}$, photon emission dominates, and  nucleons in at least the inner crust and outer core of such cold stars are thought to be superfluid. The heat capacity (per unit volume) of such very cold NS's is dominated by ultrarelativistic, degenerate electrons, and is given by \cite{Yakovlev:1999sk}
\beq
c_{V,e} \approx {p_{\text{F} e}^2 k_\text{B}^2 T \over 3} = {(3 \pi^2 n_e)^{2/3} k_\text{B}^2 T \over 3}.
\eeq 

Given heating only through DM kinetic heating, and black-body-radiation-dominated cooling, the NS temperature evolves in time according to 
\beq
C_V {d T \over d t}=\dot E_\text{kin} - 4 \pi R^2 \sigma_B T^4.
\eeq
Without a heat source, using the relations above and approximating the NS as roughly uniform temperature throughout, we approximate the timescale to cool from a higher temperature $T_h$ to a lower temperature of order $T_l$ below $10^6 \text{K}$ as 
\beq
t_\text{c}(T_l) \sim  {R (3 \pi^2 { n_e})^{2/3} k_\text{B}^2 \over 18 \sigma_B T_l^2}\left(1 - \left({T_l \over T_h}\right)^{2}\right) \sim 10^4 \yr \left( 10^4 \text{K} \over T_l\right)^2 \left(1 - \left({T_l \over T_h}\right)^{2}\right) \label{eq: cooling timescale}
\eeq
where we set the electron density to a typical value in the NS outer core, $ n_e  \sim 0.01 \text{fm}^{-3}$. To set a kinetic heating limit based on a maximum temperature, $T$, we require a deposit rate (flux) comparable to or greater than one per cooling time, $\dot N \gtrsim 1/t_c(T)$.

Going the other way, the temperature, $T_h$, to which a NS is heated from temperature $T_l$ through an energy deposit $\Delta E$ is given by 
\beq
\Delta E = \int C_V dT \sim {4 \pi R^3 \over 3 }{(3 \pi^2  n_e)^{2/3} k_\text{B}^2 \over 6}(T_h^2-T_l^2) \sim 5 \times 10^{-14} {G M^2 \over R}{T_h^2 - T_l^2 \over (10^6 \text{K})^2}
\eeq where the last expressions apply when both temperatures are below $10^6~\text{K}$. The heat capacity rises by a factor of 20 or so at higher temperatures, and we can see that $\Delta E \gtrsim  10^{-2} {G M^2 \over R}$, approaching the gravitational binding energy of a NS, roughly heats a NS back above its temperature at birth ($\sim 10^{11} ~\text{K}$). We expect a single energy deposit of order $10^{-2} {G M^2 \over R}$ or greater to destroy a NS. So the mere existence of a cold NS of age $t_\text{NS} > \dot N_\text{cap}^{-1}$ limits
\beq
{\Delta E_\text{graze} \over 10^{-2} G M^2/R} \sim 10^{4} \left({m_X \over M}\right)^2(1+\alf)^2 \sqrt{T \over 4.2 \text{kK}} < {E_\text{DM, kin}(R) \over 10^{-2} G M^2/R} \sim 10^2 {m_X \over M}(1+\alf)< 1. \label{eq: NS destruction} 
\eeq

\subsection{Limits}\label{subsec: limits}

\begin{figure}
\includegraphics[width=0.95\textwidth]{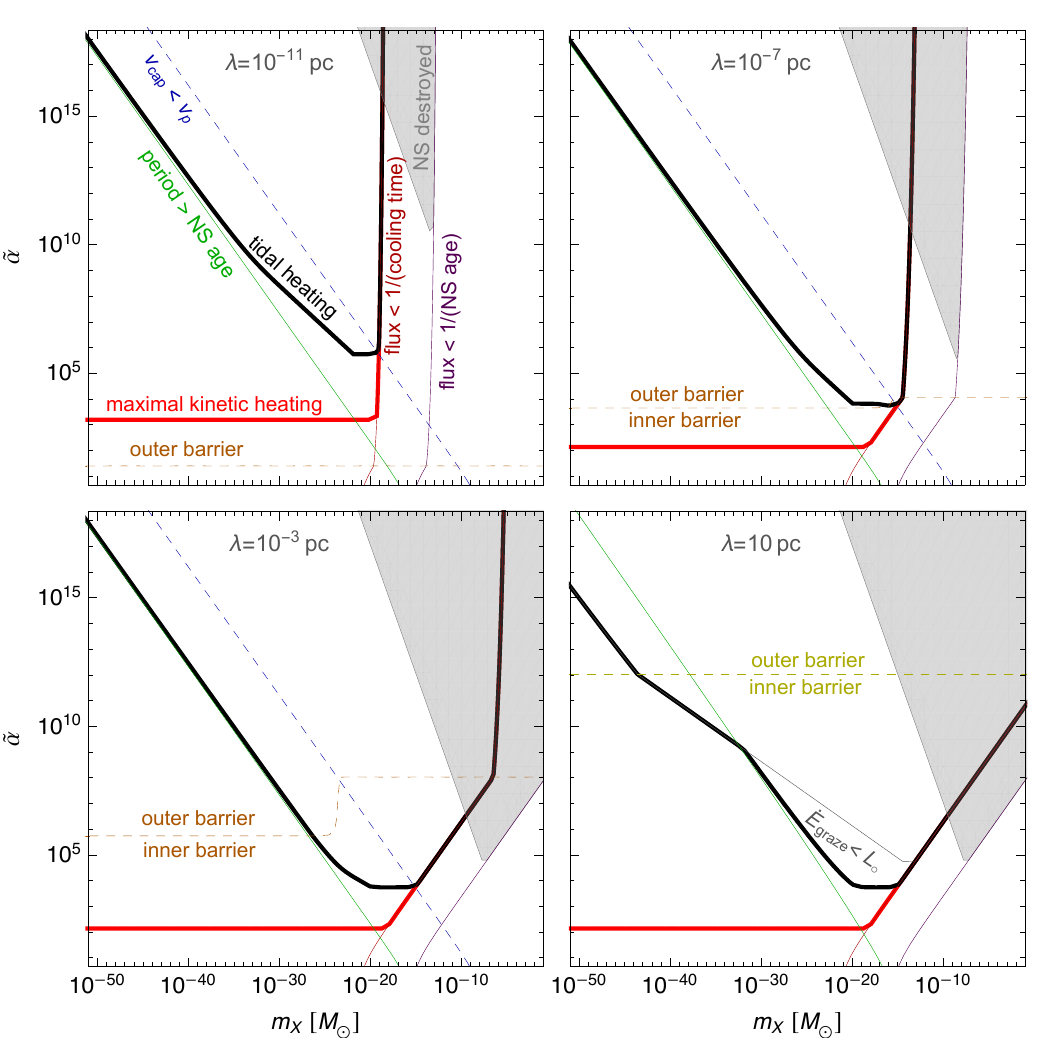}
\caption{Limit on an attractive NS-DM fifth force (\Eq{eq: yukawa}) for four representative force ranges, $\lambda$, from \emph{tidal} kinetic \emph{heating} of the coldest known NS with temperature $T < 4.2 \text{kK}$ and age $0.3 \Gyr$ (black). In \Sec{subsec: limits} we explicitly explain how limits were derived in several regimes, beginning at high mass. Eqs.~\ref{eq: flux boundary}, \ref{eq: cap flux boundary}, \ref{eq: cap heating limit approx}, and \ref{eq: delta t contour} determine the limits in the first three panels from high to low mass, respectively, and \Eq{eq: graze heating limit approx} applies at the lowest masses in the last panel. The thick red \emph{maximal kinetic heating} curve assumes 100\% of transiting DM is captured and deposits energy quickly through additional short-range forces, leading to \Eq{eq: f=1 limit approx} at low masses. In the gray regions, the old NS would have been destroyed by a single DM encounter (see \Eq{eq: NS destruction}).}\label{fig: kinetic heating panel}
\end{figure}

Here we present limits based on \Eq{eq: limit} for the coldest presently known NS \cite{Guillot_2019},  a slow, isolated pulsar located about $170 \pc$ from Earth, with the observation upper limit on temperature ${T_{\text{s}, i}}_\text{max}^{\text{meas}} = 42,000\K$ and age $t_\text{NS} = 0.3 \Gyr$. Without additional sources of heating, this NS should have cooled to temperatures below 1000~\text{K} within the first 10\% of its life. Here we emphasize that the value ${T_{\text{s}, i}}_\text{max}^{\text{meas}} = 42,000\K$ is an upper limit based on telescope observations, and is expected to tighten as observational techniques improve. The four panels in \Fig{fig: kinetic heating panel} show limits for four different force ranges, $\lambda$, spanning our range of interest. The thick black \emph{tidal heating} limit lines result from our estimate of  $\dot E_\text{kin}$ in \Eq{eq: Ekin tidal}, as described in Secs.~\ref{subsec: graze}-\ref{subsec: captured}. The thick red \emph{maximal kinetic heating} line results when an order one fraction of NS-transiting DM deposit their entire available energy over a NS lifetime, with a heating rate as in \Eq{eq: Ekinmax}, requiring assistance from short-range forces as described in \Sec{subsec: max capture}. We employed our analytic estimate of the impact parameter in Eqs.~\ref{eq: b inner}-\ref{eq: b} to calculate the relevant fluxes (Eqs.~\ref{eq: N dot graze} and \ref{eq: cap N dot}), and the limits assume asymptotic DM density $\rhoDM = 0.4 \GeV/\cm^3$, a Maxwellian speed distribution peaked at $v_p \sim 10^{-3}$, and typical NS parameters ${G M \over R} \approx 0.2$, $R \approx 10 \km$. Below we explain the scaling of the limits, starting with the high-mass behavior and working to lower masses.

\def\tcf{\bar{t}_\text{c}}
\def\rhof{\bar{\rho}_X}
\def\rf{\bar{R}}
\def\vpf{\bar{v}_p}
\def\gmf{\overline{G M}}
\def\Mf{\bar{M}}
\def\Tf{\bar{T}}

To set a limit based on \Eq{eq: limit} for a given maximum NS temperature $T$, we require the frequency of energy deposit events, $\dot N$, to be comparable to or greater than the inverse cooling timescale, $1/t_\text{c}(T)$, in \Eq{eq: cooling timescale}. This requirement determines the high-mass cut-off for the limit, shown by the thin purple ``flux $<$1/(cooling time)'' lines in \Fig{fig: kinetic heating panel}. Using \Eq{eq: N dot graze}, and assuming a Maxwellian speed distribution with peak speed $v_p$, the high-mass limit reads,
\begin{align}
\dot N_{\mathrm{graze}} t_\text{c} \sim {6 \times 10^{-22}} \left(\frac{t_c}{500\,\rm{yr}}\right)\left(\frac{R}{10\,\rm{km}}\right)^2\left(\frac{10^{-3}}{v_p}\right)\left(\frac{M_{\odot}}{m_X}\right) \left({(b_\text{max} v)|_{v_p} \over R}\right)^2
 < 1 . \label{eq: flux boundary}
\end{align}
 where fiducial quantities in the parentheses are used to obtain our \Fig{fig: kinetic heating panel} and we have taken here (and throughout) $\rhoDM = 0.4 \mbox{ GeV/cm}^3$. Note that $t_c(T_l=42\,\rm{kK})\sim 500$ years according to \Eq{eq: cooling timescale}. Here, $b_\text{max}$ is given in Eqs.~\ref{eq: b inner}-\ref{eq: b} and \Fig{fig: bmax}, and its scaling with $\alf$ depends on whether it is determined by the inner or outer barrier, as indicated by the brown horizontal dashed lines in the figure.\footnote{In the first three panels, we show the transition for $b_\text{max}|_{v_\text{cap}}$ and in the last the dashed line corresponds to the transition for $b_\text{max}|_{v_p}$.}  \Eq{eq: bmax large alpha approx} gives the scaling of $b_\text{max}$ in various regimes.  

The grazing and capture fluxes are approximately equal at the highest constrained masses, where $v_\text{cap}  \sim 5(1 + \alf)\left(\frac{m_X}{M_{\odot}}\right)^{1/2}\left(\frac{T}{42\,\rm{kK}}\right)^{1/4}\left(\frac{10\,\rm{km}}{R}\right)^{1/2} \gtrsim  v_p$ (see Eqs.~\ref{eq: dE}, \ref{eq: lmax}, and \ref{eq: vmaxcap}). The tidal heating limit curves break away from the $\dot N_\text{graze} = 1/t_\text{c}$ curves where $v_\text{cap} < v_p$, below the dashed blue lines in \Fig{fig: kinetic heating panel}. When $v_\text{cap} < v_p$, the flux-limited limit follows
\begin{align}
\dot N_\text{cap} t_\text{c} 
\sim {10^{-14}}(1 + \alf)^2\left(\frac{t_c}{500\,\rm{yr}}\right) \left(\frac{R}{10\,\rm{km}}\right)\left(\frac{T}{42\,\rm{kK}}\right)^{1/2}\left(\frac{10^{-3}}{v_p}\right)^3 \left({(b_\text{max} v)|_{v_\text{cap}} \over R}\right)^2
 < 1 \, . \label{eq: cap flux boundary} 
\end{align} 
For lower masses, the limit applies in parameter regions with continuous heating, and when $\dot E_\text{cap} \gg \dot E_\text{graze}$ (see Eqs.~\ref{eq: Ekin tidal}, \ref{eq: dE}, \ref{eq: N dot graze}, \ref{eq: cap dE}, and \ref{eq: cap N dot}), the limit reads 
\begin{align}
{\dot E_\text{cap} \over L_\circ} 
&\sim 2 \times 10^2(1 + \tilde{\alpha})^3 \left(\frac{M}{1.4\,M_{\odot}}\right)\left(\frac{T}{42\,\rm{kK}}\right)^{1/2}\left(\frac{10\,\rm{km}}{R}\right)^2\left(\frac{10^{-3}}{v_p}\right)^3 \left({m_X \over \Msolar}\right) \left({(b_\text{max} v)|_{v_\text{cap}} \over R}\right)^2  \nonumber \\
 &<  \left({{{T_{\text{s}, i}}_\text{max}^{\text{meas}} } \over 42 \text{kK}}\right)^4, \label{eq: cap heating limit approx}
\end{align}
where $L_\circ = 4 \pi R^2 \sigma_B (42 \text{kK})^4$, and the scaling of ${(b_\text{max} v)|_{v_\text{cap}}}$ with $\alf$ again depends on whether the outer or inner barrier determines $b_\text{max}$. For \Fig{fig: kinetic heating panel}, the relevant relationships are $b_\text{max} v \sim {G M} \alf$ for the inner barrier and $b_\text{max} v \sim \sqrt{2 G M \lambda \log( \alf \log \alf )} $ for the outer. 

The thin green ``period $>$ NS age'' lines in \Fig{fig: kinetic heating panel}  are the contours $\Delta t_1 = t_\text{NS}$ with $\Delta t_1$ as in \Eq{eq: period}.  Heating from captured DM goes to zero when $\Delta t_1 > t_\text{NS}$ (c.f.~\Eq{eq: cap dE}), explaining why the limit asymptotes to this contour in the first three panels of \Fig{fig: kinetic heating panel} at low DM masses.  Using the results in \App{sec: timescale for heating}, for $\lambda \ll (\sqrt{G M } t_\text{NS})^{2/3} \sim 10 \pc \left( {t_\text{NS} \over \Gyr} \right)^{2/3}$, the  $\Delta t_1 > t_\text{NS}$ asymptote corresponds to 
\begin{align}
 t_\text{NS} \over \Delta t_1 &=  {t_\text{NS} \over GM} {\sqrt{2} \over \pi}\left({\Delta E_\text{graze} \over m_X}\right)^{3/2} \nonumber \\
&\approx 3 \times 10^{22}(1 + \alf )^{3} \left(\frac{T}{42\,\rm{kK}}\right)^{3/4}\left(\frac{1.4\,M_{\odot}}{M}\right)\left(\frac{10\,\rm{km}}{R}\right)^{3/2}\left(t_\text{NS} \over 0.3 \Gyr\right)  \left({m_X \over \Msolar}\right)^{3/2}  < 1. \label{eq: delta t contour}
\end{align}

In the last panel, at low masses the contour is instead controlled by heating from grazing encounters.  When $\dot E_\text{graze} \gg \dot E_\text{cap}$, the limit is, 
\begin{align}
{\dot E_\text{graze} \over L_\circ} 
\sim 4 \times 10^{-4}(1 + \tilde{\alpha})^2\left(\frac{T}{42\,\rm{kK}}\right)^{1/2}\left(\frac{10\,\rm{km}}{R}\right)\left(\frac{10^{-3}}{v_p}\right) \left({m_X \over \Msolar}\right) \left({(b_\text{max} v)|_{v_p} \over R}\right)^2   <  \left({{{T_{\text{s}, i}}_\text{max}^{\text{meas}} } \over 42 \text{kK}}\right)^4 \label{eq: graze heating limit approx}
\end{align}
 with $(b_\text{max} v)|_{v_p}$ as in \Eq{eq: bmax large alpha approx} for large $\alf$. The  thin gray ``$\dot E_\text{graze} < L_\circ$'' contour in the last panel corresponds to saturation of \Eq{eq: graze heating limit approx} evaluated at the fiducial parameters.

When an order one fraction of transiting DM is captured and deposits its available energy quickly, kinetic heating is ``maximal.''  When $\lambda \gg R$ and $\dot N_\text{graze} t_\text{c} > 1$, the limit reads
\begin{align}
{\dot E_\text{kin, max} \over L_\circ} \sim  6 \times 10^{-6}(1 + \alf ) \left({M \over {1.4\,M_{\odot}}}\right)\left(\frac{10\,\rm{km}}{R}\right)\left(\frac{10^{-3}}{v_p}\right)  \left({(b_\text{max} v)|_{v_p} \over R}\right)^2  <  \left({{{T_{\text{s}, i}}_\text{max}^{\text{meas}} } \over 42 \text{kK}}\right)^4. \label{eq: f=1 limit approx}
\end{align} 
To one significant figure the maximal kinetic heating limit shown in red in \Fig{fig: kinetic heating panel} corresponds to 
\begin{align}
\tilde{\alpha} < 150 \qquad &\text{when}~{\lambda \gtrsim 10^{-10} \text{pc}}, \label{eq: f=1 limit, large lam}\\
{\lambda \over R} \tilde{\alpha} \log\left(\tilde{\alpha} \log(\tilde{\alpha})\right) <  4 \times10^5 \qquad &\text{when}~~ R < \lambda \lesssim 10^{-10} \text{pc} . \label{eq: f=1 limit, small lam}
\end{align}  
In \App{app: glitches}, we show that the kinetic heating limit above also rules out the possibility of observable pulsar glitches caused by angular momentum deposits of transiting DM.

Near-future measurements with infrared telescopes such as JWST should be able to improve NS surface temperature measurements by a factor of ten or better (see \cite{Baryakhtar:2017dbj}). For smaller force ranges, tidal heating limits will not change much because they are bounded by the flux and energy deposit timescales (see Eqs.~\ref{eq: flux boundary}, \ref{eq: cap flux boundary} and \ref{eq: delta t contour}) except in a narrow mass range, where \Eq{eq: cap heating limit approx} applies. The bound for larger $\lambda$ will tighten at lower masses, according to \Eq{eq: graze heating limit approx}.  If the temperature of a cold NS were an order of magnitude lower, so ${T_{s} } \sim 4,200 \text{K}$, the maximal kinetic heating limit (\Eq{eq: f=1 limit approx}) becomes $\tilde{\alpha} \lesssim 2$ for all $\lambda \gtrsim R$. Recalling that the maximum heating rate of a typical local NS by DM focussed through gravity alone corresponds to $T_{s} \sim 2,000 \text{K}$ (see \Eq{eq: max heating rate}), we expected maximal kinetic heating to constrain a fifth force to  gravitational strength or weaker for temperatures approaching this infrared range. 

 We now highlight some important caveats. First, DM-DM interactions become important to NS capture and heating at time $t_\text{eq}$ when the effective charge of captured DM is comparable to the effective charge of baryons in the NS:\footnote{This expression assumes the charge-to-mass ratio of captured DM is similar to that of halo DM. If this were untrue, additional heat would be released (absorbed) in the capture process, leading to larger (smaller) DM charge-to-mass ratios.} $\dot N_\text{cap} t_\text{eq} g_X = {M \over m_n} g_n$.  Accelerated (decelerated) capture and heating start just before $t_\text{eq}$ given scalar (vector) mediation. Therefore scalar-mediated constraints should tighten and vector mediated constraints should loosen when 
\beq
{t_\text{NS} \over t_\text{eq}} = {m_X \over M}{\alf \over \alf_n}\dot N_\text{cap} t_\text{NS} > 1.
\eeq
When this inequality is satisfied, time dependence of capture and heating rates should be taken into account. More specifically, the capture luminosity constraint, $E_\text{DM, kin}(R) \dot N_\text{cap} \approx m_X {G M \over R}(1+ \tilde{\alpha})  \dot N_\text{cap} < L_\circ \left(T_\text{max} \over 42 \text{kK} \right)^4$, is saturated in a region of parameter space with $t_\text{eq}< t_\text{NS}$ when
\beq
{1+ \tilde{\alpha} \over \alf}  \alf_n  < {L_\circ t_\text{NS} \over   G M^2 / R } \left(T_\text{max} \over 42 \text{kK} \right)^4 \approx 4 \times 10^{-11} \left(t_\text{NS} \over 0.3 \Gyr \right) \left(T_\text{max} \over 42 \text{kK} \right)^4   .
\eeq
Since WEP tests suggest $\alf_n \lesssim 10^{-11}$ when $\lambda \gtrsim 2 R_\text{Earth}$ \cite{Wagner:2012ui,Berge:2017ovy,Fayet:2018cjy}, DM-DM interactions may thus be relevant when $\alf_n$ is significantly below this bound. Our kinetic heating limits which neglect DM-DM interactions are thus conservative in the scalar-mediated case since captured DM could further accelerate capturing more DM.  On the other hand our bounds could lift in the vector-mediated case if the DM-DM repulsive interaction was sufficiently strong compared to the attractive DM-baryon interaction. Furthermore, heating that appears to accelerate (decelerate) at a characteristic timescale depending on local DM density and NS age could be a smoking gun signal of a scalar (vector) mediated long-range force.  

We also remind the reader of restrictions necessary for consistency of our tidal heating estimate discussed in Secs.~\ref{subsec: graze}-\ref{subsec: captured}. First, the DM must be sufficiently compact to survive an NS transit and to tidally excite modes up to the cutoff determined by shear viscous damping (see \Sec{subsec: dE graze}). Second, tidal energy deposits approach the expected maximum $E_\text{DM, kin}(R)$ at approximately the same $\tilde{\alpha}, m_X$ parameter range as they approach a few percent of the NS gravitational binding energy, where we expect transiting DM to destroy rather than heat the NS (see \Eq{eq: NS destruction}). \Fig{fig: kinetic heating panel} shows this region in gray.


\section{Limits from Pulsar Timing Arrays}\label{sec: PTA}

The use of PTAs to probe DM substructure has been extensively studied~\cite{Siegel_2007, Seto_2007, Baghram:2011is,Kashiyama_2012,Clark_2015_I,Schutz_2017, Dror:2019twh, Kashiyama:2018gsh, Ramani:2020hdo}. To set upper limits on the fifth force strength, we analyse the data collected by NANOGrav~\cite{brazier2019nanograv} in their 11-year dataset~\cite{Arzoumanian_2018_data, Arzoumanian_2018_GW}. This analysis utilizes the software \texttt{enterprise}~\cite{2019ascl.soft12015E} developed by NANOGrav and closely follows the Bayesian inference framework developed in our previous work~\cite{lee2021bayesian}.

We commence with deriving the phase shift signals measured by an observer on Earth due to a transiting DM. The intrinsic pulsar phase, $\phi(t)$, is often modelled as a truncated power series in $t$
\beq
\phi(t) = \phi_0 + \nu t+ \frac{1}{2}\dot{\nu}t^2\, ,
\label{eq:timing_model}
\eeq
where $\phi_0$ is the phase offset, $\nu$ is the pulsar frequency and $\dot{\nu}$ is its first derivative. Astrophysical signals such as Doppler shifts due to transiting DM manifest as deviations from the above timing model. 

The phase shift $\delta \phi(t)$ is related to the frequency shift $\delta \nu(t)$ by $\delta\phi(t) = \int_0^t \delta \nu(t') dt'$. In the presence of an external potential $\Phi$, the acceleration of the pulsar induces an observed frequency shift due to the Doppler effect, which is given by~\cite{Dror:2019twh}
\beq
\frac{\delta \nu}{\nu} = \hat{d} \cdot \int \nabla\Phi dt\, ,
\label{eq:dnu_dop}
\eeq
where $\hat{d}$ is the unit vector pointing from Earth to the pulsar. The fifth force potential is given by Eq.~\eqref{eq: yukawa} and $\Phi(r)=V_{\mathrm{Yuk}}(r)/M$, and its gradient is
\beq
\nabla\Phi(r) = \frac{Gm_X}{r^2}\left[1+\tilde{\alpha}(1+r/\lambda)e^{-r/\lambda}\right]\hat{r} \, .
\label{eq:yukawa_gradient}
\eeq
To simplify these expressions, we perform the analysis in the pulsar rest frame and place the pulsar at the origin. We then write the DM position as $\vec{r}(t)=\vec{r}_0+\vec{v}t$ with $\vec{v}$ being the DM velocity. The two important timescales in this system are the time of closest approach, $t_0=-\vec{r}_0\cdot\vec{v}/v^2$, and the signal width,  $\tau=|\vec{r}_0\times\vec{v}|/v^2$. The impact parameter is given by $\vec{b}=\vec{r}_0+\vec{v}t_0=v\tau$. Defining the dimensionless time variable, $x\equiv(t-t_0)/\tau$, we write $\vec{r}=b(\hat{b}+x\hat{v})$ and $r=b\sqrt{1+x^2}$. Using these variables, the frequency shift due to the fifth force can be written as
\beq
\left(\frac{\delta \nu}{\nu}\right)_{\mathrm{fifth}} = \tilde{\alpha} G m_X\frac{1}{v^2\tau} \hat{d} \cdot \int \frac{1}{(1+x^2)^{3/2}}\left(1+\frac{b}{\lambda}\sqrt{1+x^2}\right)e^{-(b/\lambda)\sqrt{1+x^2}}(\hat{b}+x\hat{v})dx \, .
\label{eq:dnu_yukawa}
\eeq
The equivalent expression for gravitation is given by \Eq{eq:dnu_yukawa} but with $\tilde{\alpha}=1$ and $\lambda\to\infty$. In general the integral in \Eq{eq:dnu_yukawa} has to be computed numerically. An additional integration over time gives the phase shift $\delta\phi(t)$.

The above analysis assumes that the DM trajectory is approximately a straight line, which has been shown to hold for the DM mass range that we are interested in for gravity only~\cite{lee2021bayesian}. In the presence of a fifth force, the deviation in DM paths is small if and only if $b\approx r_{\min}$ where $r_\text{min}$ is the distance of closest approach. \Eq{eq: b of rmin} implies $b\approx r_\text{min}$ iff ${G M \over b v^2}(1 + \alf e^{-b/\lambda}) \ll 1$. One can estimate the minimum impact parameter of all passing DMs to be
\beq
b_{\min} \sim \left(\frac{3}{4\pi}\frac{m_X}{\rhoDM}\right)^{1/3}\sim 3\,\mathrm{pc}\left(\frac{m_X}{M_{\odot}}\right)^{1/3} \, .
\label{eq:bmin}
\eeq
If the aforementioned condition is not satisfied, then DMs will substantially converge to the pulsar, which will lead to a larger amplitude for the timing deviation. Hence our constraints based on the straight-trajectory approximation serve as a conservative estimate. We leave a detailed analysis of PTA constraints accounting for full DM orbit information as a potential direction for future work. 

To search for the DM signal in experimental data, similar to the static analysis in Ref.~\cite{lee2021bayesian}, we parametrize the signal with the leading order perturbation of the timing model
\beq
\frac{\delta\phi(t)}{\nu} = \frac{A}{\yr^2} t^3 \, ,
\label{eq:dphi_dop_stat}
\eeq
where $A$ characterizes the signal amplitude. Terms of order $t^2$ or less are degenerate with the timing model and have no observable consequences. We search for the signal in \Eq{eq:dphi_dop_stat} using \texttt{enterprise} and compute the Bayesian posterior distribution of the DM amplitude, $P(A|\delta\vec{t}\,)$, using the Markov Chain Monte Carlo sampling techniques with the \texttt{PTMCMCSampler} package \citep{2019ascl.soft12017E}. The parameters and priors are listed in Table.~\ref{tab:prior}. In particular, we adopt a uniform (instead of log-uniform) prior on the DM amplitude $A$, which is a standard procedure for upper-limit setting~\cite{Arzoumanian_2016, lee2021bayesian}. For the red noise amplitude, however, we use a uniform prior to avoid the transfer of signal power to the red noise process, which has been shown to lead to overstated Bayesian upper limits~\cite{Hazboun_2020}. 


\begin{table*}[t]
	\renewcommand{\arraystretch}{1}
	\begin{tabular}{cccc}
		\hline\hline
		Parameter & Description & Prior & Comments \Tstrut\Bstrut \\ \hline
		\multicolumn{4}{c}{Red noise} \\ 
		$A_{\red}$ & Red noise power-law amplitude & Log-Uniform [$-20$, $-11$]	& one parameter per pulsar \Tstrut \\
		$\gamma_{\red}$ & Red noise power-law spectral index & Uniform [0, 7]	& one parameter per pulsar \Tstrut\Bstrut \\ \hline
		\multicolumn{4}{c}{Dark Matter} \\
		$A$ & Dark matter amplitude & Uniform $\pm$[$10^{-20}$, $10^{-11}$]	& one parameter per pulsar \Tstrut \\
		\hline\hline
	\end{tabular}
	\caption{\label{tab:prior} Parameters and priors used in the PTA analysis with the 11-yr dataset from NANOGrav. The notation Uniform $\pm$[$\dots$] stands for the union of Uniform [$+\dots$] and Uniform [$-\dots$]. We account for the effects from white noise by marginalizing over a multiplicative pre-factor of the timing residual errors. Errors from Solar System ephemeris (SSE) modeling are corrected using \textsc{BayesEphem}, as described in~\cite{Arzoumanian_2018_GW}.}
\end{table*} 


We now fix a choice of $\lambda$. To relate the posterior distribution $P(A|\delta\vec{t}\,)$ to the fifth force parameters $\tilde{\alpha}$, we use a Monte Carlo simulation described in Ref.~\cite{lee2021bayesian} to compute the conditional probability $P(A|\tilde{\alpha})$. In particular, we randomly distribute DMs in a sphere and compute the total $\delta\phi(t)$ using \Eq{eq:dnu_yukawa}. We then numerically fit $\delta\phi(t)$ with a third order polynomial in $t$ to extract the  $t^3$ coefficient and hence $A$. This procedure is repeated for numerous realizations to obtain the required distribution $P(A|\tilde{\alpha})$. The final posterior distribution on $\tilde{\alpha}$ is then given by~\cite{lee2021bayesian}
\beq
P(\tilde{\alpha}|\delta \vec{t}\,) \propto \prod_{i=1}^{N_P} \int_{-\infty}^{\infty} P(A_i|\tilde{\alpha})P(A_i|\delta \vec{t}\,) dA_i
\label{eq:post_prop_multi}
\eeq
where $N_P$ is the number of pulsars.\footnote{In Ref.~\cite{lee2021bayesian} we also considered the possibility of deriving upper limits using the maximum amplitude among all pulsars (instead of using the amplitude in every single pulsar), but the limits turn out to be less stringent than the present results.} The posterior distribution satisfies the normalization condition $\int d\tilde{\alpha} P(\tilde{\alpha}|\delta \vec{t}\,)=1$.

\begin{figure}
	\includegraphics[width=0.85\textwidth]{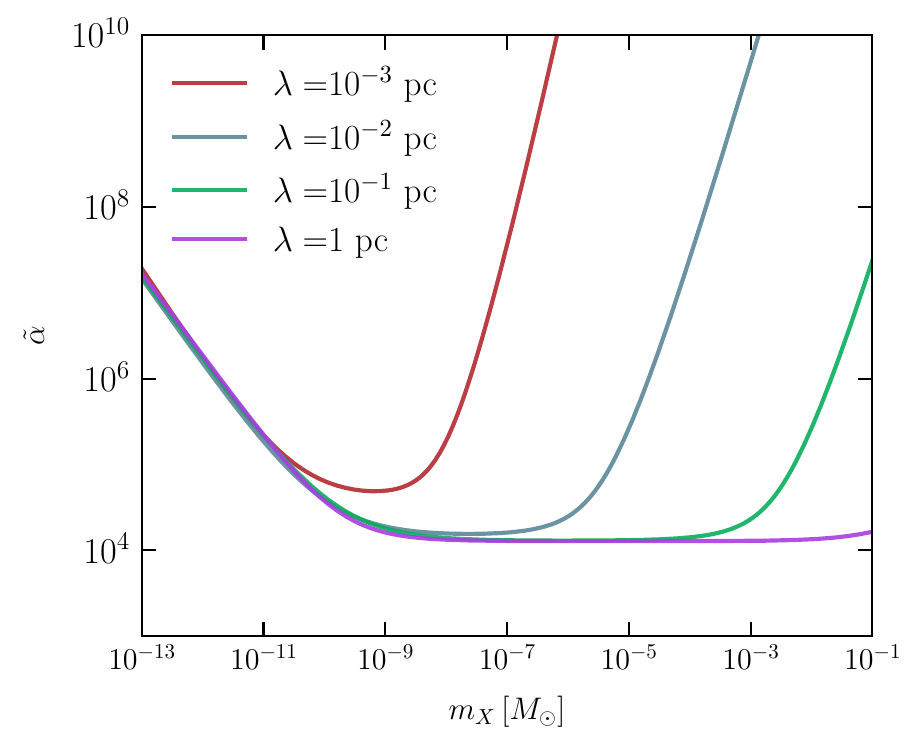}
	\caption{The 90$^\text{th}$ percentile upper limits on the fifth force strength $\tilde{\alpha}$ derived from the NANOGrav 11-year dataset. The constraints are shown for different DM mass $m_X$ and range $\lambda$.}\label{fig: pta constraints}
\end{figure}

Finally, we compute the $90^\text{th}$ percentile upper limit on $\tilde{\alpha}$ for each choice of $\lambda$, which is shown in \Fig{fig: pta constraints}. In the large mass limit, the constraints degrade since $b_{\min}$ becomes larger than $\lambda$, hence the phase shift is dominated by gravitational effects only, which is not sufficient to produce an observable signal. On the other hand, in the low mass limit, we have $b_{\min}\lesssim \lambda$, but the constraints also become less stringent, since the fifth force and gravitational strength weaken with the DM mass. For adequately large force ranges (i.e. $\lambda\gtrsim 10^{-2}$ pc), there is an intermediate mass regime where $b_{\min}\lesssim \lambda$ so the fifth force effectively modifies the gravitational constant by $G\to(1+\tilde{\alpha})G$, but its strength is not enough to perturb the DM path, hence the constraints exhibit a plateau behavior similar to the gravity-only analyses in Ref.~\cite{lee2021bayesian}. 

\section{Inferred Limits from Equivalence Principle Tests and the Bullet Cluster}\label{sec: wep bc}

In this section we consider indirect constraints on a fifth force between NS matter and DM from a combination of weak equivalence principle (WEP) tests \cite{Adelberger_2003,Wagner:2012ui}, which can constrain composition-dependent forces to be weaker than gravitational at macroscopic scales, and the bullet cluster bound on DM self-interactions \cite{Spergel:1999mh, Kahlhoefer:2013dca}. In \Sec{subsec:ep} we translate a WEP bound on the differential acceleration of two baryonic bodies toward our Galaxy's center to a limit on $\alf$ and $\lambda$ when $\lambda \ll \kpc$. We then combine bullet cluster limits on $\alf_X = {g_X^2 / 4 \pi  G m_X^2}$ with the very strong WEP constraints on $\alf_n = {g_n^2 / 4 \pi G m_n^2}$ at Earth and solar system scales ($\lambda \gtrsim 1000 \km$) to infer a limit on $\alf = \sqrt{\alf_X \alf_n}$ in \Sec{subsec:sidm}. Our results are shown in \Fig{fig:charge per mass constraint}.

\subsection{Weak Equivalence Principle Tests }\label{subsec:ep}

Most DM-baryonic matter WEP tests constrain the difference in acceleration between two baryonic test bodies (which we will label as $A$ and $B$) toward the galactic center, divided by the total acceleration:\footnote{See Ref.~\cite{Wagner:2012ui} for one of the most highly cited treatments/reviews from the last decade.}
$
 {\Delta a \over a_\text{tot}}
$.
 When the total acceleration is dominated by gravity and the fifth force is Coulombic, assuming the DM charge-to-mass ratio is constant for DM throughout the halo, it is given by\footnote{We neglect the difference between the atomic mass unit, $u$, and the neutron mass, $m_n$, throughout.} 
\beq
{\Delta a_\text{Coul} \over a_\text{tot}} = \alf \left[ \left( {Q \over M/m_n}\right)_{A}  - \left( Q \over M/\mn \right)_{B} \right], \label{eq: eta param}
\eeq where $Q$ is the effective charge of the body in units of $g_n$.
The difference in charge-to-mass ratio for the test bodies must be nonzero in order to obtain a non-trivial constraint. For normal matter, we consider couplings to baryon number or to $B-L$---i.e.~$Q = B$ or $B-L$, but other combinations lead to similar constraints.  \\ \\

\begin{table}
\begin{tabular}{| l l l| c | c |}
\hline
limit & ref & test  (test objects) & $ \Delta((B-L)/(M/m_n))$ & $\Delta(B/(M/m_n))$ \\
\hline
$|{\Delta a \over a_\text{tot}}| \lesssim 3 \times 10^{-4}$ & \cite{Sun_2019} & perihelion precession (Sun-Mercury) & 0.37 & $\sim 0$ \\
$|{\Delta a \over a_\text{tot}}| \lesssim 0.004 $ & \cite{Shao:2018klg}  & binary pulsar (NS-WD) & 0.69 & 0.19 \\
$|{\Delta a \over a_\text{tot}}| \lesssim 10^{-5} $ & see \cite{Wagner:2012ui}  & lunar laser ranging (Earth-Moon) & 0.012 & $ \sim 0$ \\
 $ |{\Delta a \over a_\text{tot}}| \lesssim 10^{-4} $& \cite{Wagner:2012ui} & torsion pendula (Be-Ti, Be-Al) & 0.013, 0.036 & 0.0024, 0.0020 \\
\hline
\end{tabular}
\caption{Collection of limits on the differential acceleration of two baryonic test objects toward the galactic center. The $B$ and $B-L$ number to mass ratios are based on Table II in \cite{Sun_2019} and Table 1 in \cite{Wagner:2012ui}. The magnitude of the gravitational acceleration field due to galactic dark matter at Earth was taken as $5 \times 10^{-11} \m/\s^2$ in \cite{Wagner:2012ui}, following \cite{PhysRevLett.70.119}; this is the value assumed in calculating $|{\Delta a \over a_\text{tot}}|$ for the last two entries.   }\label{tab: equiv principle constraint table}
\end{table}
 
 \def\rp{{\vec r}{\,'}}
 \def\rdiff{\Delta{r}}

To include range dependence in the differential acceleration calculation, we integrate over the force times DM distribution and correct \Eq{eq: eta param} with the ratio
$
 {|\Delta \vec{a}_\text{Yuk}(\vec r)| \over |\Delta \vec{a}_\text{Coul}(\vec r)| }
$. 
Given a spherically symmetric DM halo mass distribution, $\rho_{\rm{halo}}(r')$, and a uniform DM charge-to-mass ratio, the net acceleration due to a Coulombic force from halo DM on object $i$ at position $\vec r$ relative to the halo's center  is $ \vec{a}_\text{Coul}(\vec{r}) =  \alf G {Q_i \over M_i/m_n} {\hat r \over r^2} \int_0^r \rho_{\rm{halo}}({r}')  d^3 \rp $. For a Yukawa force, when $\lambda \ll r$, we find\footnote{We dropped terms suppressed by $e^{-r/\lambda}$ and higher powers of $\lambda \over r$. More precisely, we assumed $\lambda \rho_{\rm{halo}}'(r) \gg e^{-r/\lambda}\rho_{\rm{halo}}(r), {e^{-r/\lambda} \over r \lambda }\int e^{-r'/\lambda} \rho_{\rm{halo}}(r') \, r' d r', {\lambda \over r} \lambda^2 \rho_{\rm{halo}}''(r), \ldots$}
\begin{equation}\label{eqn:yukawa_final}
	\vec{a}_{\text{Yuk}}(\vec{r}) =  -\alf G {Q_i \over M_i/m_n}  \vec \nabla\int_\text{DM halo}\frac{\rho_{\rm{halo}}(r')e^{-|\vec{r}-{\vec r}{\,'}|/\lambda}}{|\vec{r}-{\vec r}{\,'}|}d^3{\vec r}{\,'}  \approx -4\pi \alf G \lambda^2{Q_i \over M_i/m_n}  \rho_{\rm{halo}}'(r)\hat{r}  \, .
\end{equation} 
Therefore when $\lambda \ll r$ the ratio of Yukawa to Coulomb differential accelerations for test objects near the same radial distance, $r$, from the halo center is given by
\beq
{|\Delta \vec{a}_\text{Yuk}(\vec r)| \over |\Delta \vec{a}_\text{Coul}(\vec r)| } \approx {- 4 \pi r^4 \rho_{\rm{halo}}'(r) \over \int_0^r \rho_{\rm{halo}}(r') d^3 \rp} \left(\lambda \over r \right)^2. \label{eq: force ratio est}
\eeq
Given an NFW DM halo profile \cite{Navarro:1995iw}, 
\beq
	\rho_{\rm{halo}}(r) = \frac{\rho_{\rm{halo},0}}{r/r_s(1+r/r_s)^2} ,
\eeq
with scale radius $r_s$, the expression in Eq.~\eqref{eq: force ratio est} becomes
\beq
{|\Delta \vec{a}_\text{Yuk}(\vec r)| \over |\Delta \vec{a}_\text{Coul}(\vec r)| } = \frac{(r/r_s)^2(1+3r/r_s)}{(1+r/r_s)^2\left[(1+r/r_s)\log(1+r/r_s)-r/r_s\right]}\left(\frac{\lambda}{r}\right)^2. \label{eq: force ratio NFW}
\eeq
The coefficient in Eq.~\eqref{eq: force ratio NFW} lies between $2.4$ and $2.6$ when $0.23 < {r \over r_s} < 2$, with the maximum of about $2.6$ occurring near ${r \over r_s} = 0.8$. For concreteness we take $r = 8 \kpc$ and set the coefficient to $2.5$, which is within 4\% of the exact value when $4 \kpc  < r_s < 35 \kpc$, well within agreement with recent fits \cite{Cautun:2019eaf,Read:2014qva}. The constraint on a Yukawa fifth-force then reads,
\beq
\left|{\Delta a_\text{Yuk} \over a_\text{tot} }\right|  \approx 2.5 \, \alf  \left({\lambda \over r} \right)^2  \left[ \left( {Q \over M/m_n}\right)_{A}  - \left( Q \over M/\mn \right)_{B} \right] <  \left | {\Delta a \over a_\text{tot}} \right|^\text{max} .
\eeq 
 \Tab{tab: equiv principle constraint table} shows a variety of recent limits on differential accelerations, along with the difference in $B$- and $(B-L)$-to-mass ratios for the test objects. The strongest limit given $Q = B$ corresponds to $|{\Delta a \over a_\text{tot}}| / |\Delta[ Q/(M/m_n)]| \lesssim 0.02$. The best limit given $Q = B-L$ is more than an order of magnitude stronger. Therefore the limit on a Yukawa fifth force coupling to baryon number from equivalence principle tests is roughly,
\beq
2.5 \, \alf \left( { \lambda \over 8 \kpc} \right)^2 \lesssim 0.02 \qquad \text{when}~\left({m_X \over \rhoDM}\right)^{1/3} \ll \lambda \ll 8 \kpc \label{eq: equiv limit},
\eeq
where the conditions on $\lambda$ are necessary for consistency of our spherically symmetric fluid approximation for the DM distribution near Earth, where the inter-DM spacing is estimated as $\left({m_X \over \rhoDM}\right)^{1/3}$ and we set Earth's galactic radius to $r\approx8 \kpc$.

\subsection{Bullet Cluster limit}\label{subsec:sidm}

Following Ref.~\cite{Coskuner:2018are}, the momentum transfer cross section for a long-range  DM-DM interaction can be approximated as $\sigma_T \approx 4 \pi d_C^2$ where $d_C$ is the distance of closest approach given a repulsive Yukawa interaction, $V_\text{Yuk} = {G m_X^2 {\tilde{\alpha}_X} \over r} e^{-r/\lambda}$. For simplicity we assume an order one fraction of the DM by mass takes the form of objects with roughly the same size. Assuming the Yukawa DM-DM  interaction is much stronger than the gravitational DM-DM interaction at closest approach, $ {d_C \over \lambda}  = W\left({4 G \tilde{\alpha}_X m_X  \over \lambda v^2}\right)$, where $W$ is the Lambert $W$ function satisfying $W(z) e^{W(z)} = z$ and $v$ is the relative speed of the interacting DM pair. Note that $\sigma_T$ thus depends on the DM velocity $v$. Observation of the bullet cluster sets the rough limit ${\sigma_T \over m_X} = {4 \pi d_C^2 \over m_X} < {\cm^2 \over \text{g}}$, translating to a limit 
\beq
{4 G \tilde{\alpha}_X m_X  \over v^2} < \sqrt{{1 \over 4 \pi}{m_X \over \text{g}}} e^{\sqrt{{1 \over 4 \pi}{m_X \over \text{g}}}{ \cm \over \lambda}} \cm \label{transfer xsec}
\eeq or 
\beq
 \tilde{\alpha}_X  \sqrt{m_X \over \Msolar} < 2 \times 10^6 \left({v \over 10^{-2}}\right)^2 e^{4 \times 10^{-3}\sqrt{{m_X \over \Msolar}}{ \pc \over \lambda}}. \label{eq: bullet limit}
\eeq
The limit becomes very weak when the interaction range, $\lambda$, is comparable or small compared to the scattering length $\sqrt{m_X \over \text{g}} \cm = 10^{-2} \sqrt{{m_X \over M_\odot}} {\pc} $.\footnote{Conversely, the limit is very insensitive to $\lambda$ as long as $\lambda \gg 10^{-2} \sqrt{{m_X \over M_\odot}} {\pc} $.} 
The constraint on $\tilde{\alpha}_X$ is clearly consistent with our assumption that $\tilde{\alpha}_X \gg 1$ when $m_X \lesssim M_\odot$.

Now consider the indirect constraint on $\alf$.  If the DM and B fifth force charge to mass ratios are approximately independent of environment, then
\beq
\tilde{\alpha} \approx \sqrt{\tilde{\alpha}_n \tilde{\alpha}_X}.
\eeq
Therefore the bullet cluster constraint plus a constraint on $\alf_n$ lead to a constraint on $\alf$:
\beq
\tilde{\alpha} < 5 \times 10^{-3} \left({{\tilde{\alpha}}_n^\text{max} \over 10^{-11} }\right)^{1/2} \left( \Msolar \over m_X \right)^{1/4}  \left( v \over 10^{-2} \right) e^{2 \times 10^{-3} \sqrt{{m_X \over \Msolar}}{\pc \over \lambda}}.\label{eq: inferred}
\eeq
We choose a fiducial value of $10^{-11}$ for $\tilde{\alpha}_n^\text{max}$ based on the MICROSCOPE limit \cite{Berge:2017ovy,Fayet:2018cjy} but note that a more conservative limit based on inverse-square-law tests corresponds to $\tilde{\alpha}_n^\text{max} \sim 10^{-1}$ - $10^{-10}$ in the $\km$ to $\pc$ force range \cite{Adelberger_2003}. Furthermore, both the inverse square law and MICROSCOPE limits come from measuring accelerations on Earth and in our solar system---very different from the bullet cluster environment. A more conservative approach would not combine the limits at all. Furthermore, we note that the bullet cluster limit on DM sub-components weakens precipitously.

\begin{figure}
\includegraphics[scale=1]{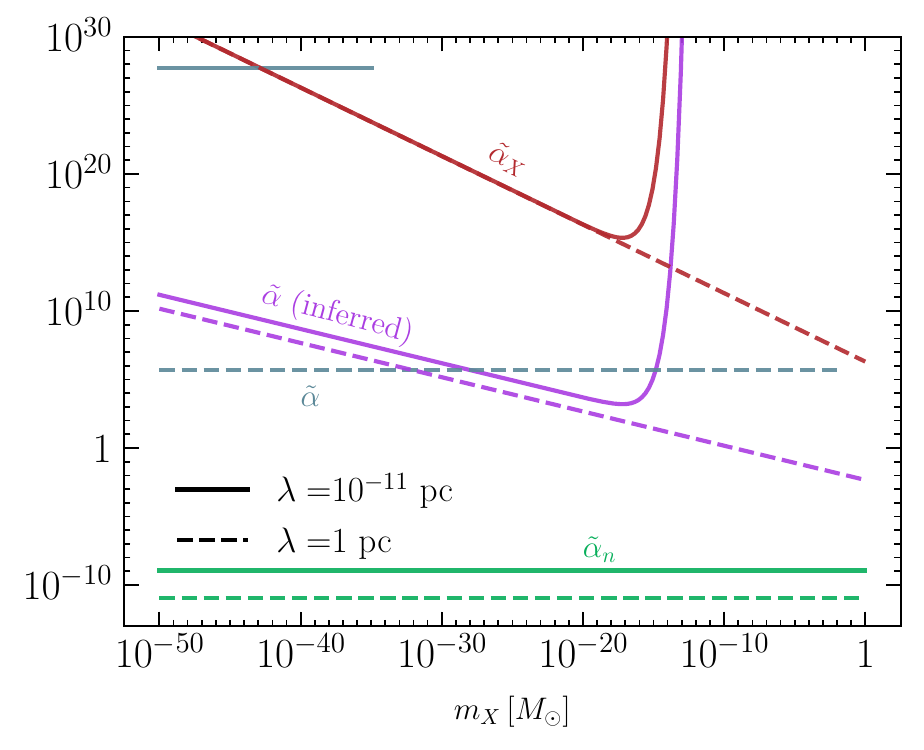} 
\caption{Constraints on the fifth force to gravitational force strength ratios for baryonic matter self-interactions ($\alf_n$), DM-baryon interactions ($\alf$), and DM self-interactions ($\alf_X$), from MICROSCOPE  (Fig.~1 of \cite{Berge:2017ovy}), weak equivalence principle tests constraining accelerations toward the galactic center (\Eq{eq: equiv limit} and \Tab{tab: equiv principle constraint table}), and the bullet cluster (\Eq{eq: bullet limit}), respectively, for force ranges at the extremes of those we consider. The inferred constraint on $\alf$ (purple) is derived by combining the MICROSCOPE (green) and bullet cluster (red) limits, using $\tilde{\alpha}^\text{max} = \sqrt{\tilde{\alpha}_n^\text{max} \tilde{\alpha}_X^\text{max}}$ (\Eq{eq: inferred}).  We have assumed that all of the DM in the Milky Way halo and in the bullet cluster takes the form of compact, effectively point-like states with approximately the same mass, $m_X$.}\label{fig:charge per mass constraint}
\end{figure}

The limits in Eqs.~\ref{eq: bullet limit} (red), \ref{eq: inferred} (purple), and \ref{eq: equiv limit} (blue) are represented in \Fig{fig:charge per mass constraint}, along with the best constraint on $\tilde{\alpha}_n$, from Fig.~1 of \cite{Berge:2017ovy} (green) for two force ranges at the boundaries of our range of interest. The inferred limit on $\alf$ is just the geometric mean of the limits on $\alf_X$ and $\alf_n$.

\section{Summary and Conclusions}\label{sec: conclusion}

We investigated probes of a long-range DM-baryon interaction. First, we derived limits on the interaction strength and range as functions of NS temperature and DM mass, density, and asymptotic speed distribution. This work extends the dark kinetic heating analysis of \cite{Baryakhtar:2017dbj} to long-range forces and draws on the PBH gravitational tidal capture analysis of \cite{2014}.  \Fig{fig: kinetic heating panel} shows our limit based on the coldest known NS. Next, extending the work in \cite{lee2021bayesian}, we considered effects of a long-range fifth force on Doppler shifts of pulsar frequencies, and derived a limit based on the 11-yr NANOGrav PTA timing dataset, shown in \Fig{fig: pta constraints}. Finally, we considered \emph{indirect} limits from weak equivalence principle tests and the bullet cluster, shown in \Fig{fig:charge per mass constraint}. The three sets of constraints are shown together in \Fig{fig: money}. The indirect  bullet cluster $+$ WEP bound is stronger than the direct tidal heating and PTA bounds by an order of magnitude or more for all of the parameter space we considered. However, the tidal heating and PTA constraints are still interesting since they are direct phenomenological probes on DM-baryon interactions, independent of the microscopic origin of such a force. Moreover, if additional short-range interactions assist DM capture by NSs (allowing ``maximal kinetic heating'') or  if only a subcomponent of DM interacts through the long-range fifth force, the kinetic heating and PTA limits dominate. 

Imminent improvements on NS temperature and pulsar timing observations from, e.g., the James Webb Space Telescope \cite{Gardner:2006ky} and Square Kilometer Array \cite{Weltman:2018zrl} will extend the reach of NSs as probes of a long-range DM-baryon fifth force. In addition, the requirement for the NS cooling timescale to be smaller than the encounter timescale can be potentially relaxed by surveying a large population of NSs and searching for a fraction of them that are still cooling down after recent DM encounters~\cite{Bramante:2021dyx}. On the other hand, higher than expected temperatures of isolated old NSs could be a sign of DM kinetic heating; if those temperatures are higher than can be explained given kinetic heating through short-range interactions (c.f. \cite{Baryakhtar:2017dbj} and \Eq{eq: Ekinmax} with $\alf = 0$), a long-range force could be part of the explanation. So could DM annihilation (see e.g.~\cite{Kouvaris:2007ay,Bertone:2007ae,Kouvaris:2010vv,deLavallaz:2010wp,Bramante:2017xlb}) or other DM-induced exothermic processes (see e.g.~\cite{McKeen:2020oyr, McKeen:2021jbh}). An unexpected NS temperature age dependence for old NSs in otherwise similar environments is another potential signature of a long-range force, since the heating rate can dramatically change once the effective charge of captured DM becomes comparable to that of the NS at birth. As discussed at the end of \Sec{sec: kinetic heating}, scalar-mediated interactions would lead to accelerated heating and vector-mediated to decelerated heating at late times.  We leave detailed analyses of such possibilities for future work, should any observational hints arise.

\acknowledgments

We thank Stephen Taylor and the NANOGrav collaboration for providing valuable guidance in deriving the PTA constraints. Some of this work was performed at the Aspen Center for Physics, which is supported by National Science Foundation grant PHY-1607611. MG's work was also supported by the National Science Foundation under Grant No.~1719780. The work of VL and KZ is supported by the DoE under contract DE-SC0011632, and by a Simons Investigator award. This work is also supported by the Walker Burke Institute for Theoretical Physics. 


\appendix


 \section{Classical Orbits and $b_\text{max}$}\label{app: orbits}

In this appendix we consider orbits of point-like DM about a static NS given both a gravitational and Yukawa force between them, as described in Sec.~\ref{sec: introduction}. Our primary goal is to identify the impact parameters of orbits that intersect the NS, which requires finding the location of centrifugal barriers. The barriers can occur at radii $r \gtrsim \lambda$ where the fifth force is starting to turn on, and at radii much smaller than the force range, where general and special relativity can be relevant. We begin with a reminder of the gravity-only case and then generalize to include the Yukawa interaction.

The  general relativistic expression of energy conservation given a spherically symmetric star of mass $M$ and a much lighter orbiting body of mass $m$ comes from the constraint $p_\mu p^\mu = - m^2$. Employing the Schwarzschild metric and coordinates, $p_t = g_{t t} m {d t \over d \tau} = - E = - m \gamma$ and $p_\phi = g_{\phi \phi} m {d \phi \over d \tau} = L = \gamma m b v$ are conserved quantities along geodesics, where $v$ is the asymptotic DM speed, $\gamma = 1/\sqrt{1 - v^2}$, and $b$ is the impact parameter. With these identifications the constraint reads,
\beq
\left({d r \over d \tau}\right)^2 = {E^2 \over m^2} -  \left({(L/m)^2 \over r^2}+1 \right)\left(1 - {2 G M \over r} \right) . \label{four momentum scalar product}
\eeq
Given a large enough impact parameter, DM streaming in from infinity with asymptotic speed $v$ hits a centrifugal barrier, where ${d r \over d \tau} \rightarrow 0$. 

Given an additional attractive Yukawa interaction, the four-momentum constraint is the same, but $E \rightarrow E - V_\text{Yuk}$. So orbits obey\footnote{The NS's fifth force charge could non-negligibly affect the general relativistic lapse function when ${G \af Q_\text{NS}^2 \over R^2} = \alf_n \left({\left(Q \over M/m_n\right)_\text{NS}} {G M \over R}\right)^2 \gtrsim 0.1$. So a consistency requirement is $\sqrt{\alf_n} \left( Q \over M/m_n \right)_\text{NS} \ll  1$. Since equivalence principle measurements restrict $\alf_n \lesssim 10^{-11} \text{-} 10^{-9}$ for the force ranges we examine, the condition easily holds when baryonic charge (and not captured DM charge) dominates the total NS charge.} 
\beq
\left({d r \over d \tau}\right)^2 = \left({E-V_{\text{eff}, +}  \over m} \right)  \left({E-V_{\text{eff}, -}  \over m} \right) \label{eq: orbit eom}
\eeq
with\footnote{C.f.~\cite{Olivares:2011xb}.}
\beq
{V_{\text{eff}, \pm} \over m} = -{G M \tilde{\alpha} e^{-r/\lambda} \over r} \pm \sqrt{\left(1 - {2 G M \over r} \right)\left({(L/m)^2 \over r^2} +1\right) }. \label{eq: veff rel yukawa}
\eeq

The turning points, $r_\text{min}$, of unbound orbits---i.e. centrifugal barriers---occur at the maximum radial coordinate for which $V_{\text{eff}, +} = E$. Solving the condition for the impact parameter, $b$, yields,
\beq
b(r_\text{min}) = r_\text{min} \sqrt{{1+{2 G M \over r_\text{min} \gamma^2 v^2} \left( 1 + \alf e^{-r_\text{min}/\lambda}\right) +\left({G M \over r_\text{min} \gamma v} \alf e^{-r_\text{min}/\lambda}\right)^2 \over \left(1 - {2 G M \over r_\text{min}} \right)}}.
 \label{eq: b of rmin}
\eeq
DM focused through gravity alone breaches a NS surface iff $b < b(R)|_{\alf \rightarrow 0} = R \sqrt{1 + {{2 G M / R v^2}\over 1 - 2 G M / R}} \approx \sqrt{{{2 G M R/ v^2}\over 1 - 2 G M / R}} $.

\begin{figure}
\includegraphics[width=1 \textwidth]{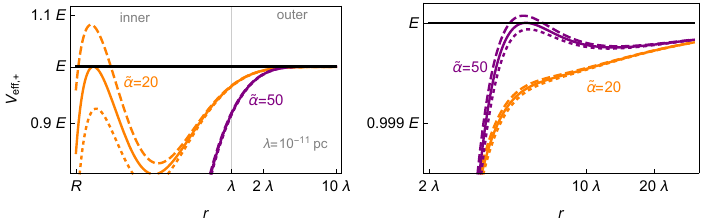} 
\caption{$V_{\text{eff}, +}$ as a function of DM radial coordinate when $b = b_\text{max}$ (solid) and $b = 10^{\pm 0.01}b_\text{max}$ (dashed/dotted), with $\lambda=10^{-11}\pc$, $v=10^{-3}$, and $\tilde{\alpha}=20$ (orange) or $\tilde{\alpha}=50$ (purple).  The right-hand plot shows only the outer region, with the vertical (logarithmic) scale magnified. In this example, the inner barrier determines $b_\text{max}$ for $\tilde \alpha = 20$ while the outer barrier determines $b_\text{max}$ for $\tilde \alpha = 50$. C.f.~\Fig{fig: bmax}.}\label{fig: veff examples}
\end{figure}

For large $\tilde \alpha$, the exponential turn-on of the fifth force can lead to an additional partial centrifugal barrier at $r > \lambda$, and the peak of the inner centrifugal barrier near the Schwarzschild radius moves to larger radial coordinates. These behaviors are demonstrated in Fig.~\ref{fig: veff examples}, which shows examples of $V_\text{eff, +}$ and $E$ assuming $v =10^{-3}$ for $\lambda = 10^{-11} \pc$. When $ \tilde{\alpha}=20$ (orange), the centrifugal barrier appears first in the inner region where $r < \lambda$. When $\tilde\alpha = 50$ (purple), the centrifugal barrier appears in the outer region where $r > 2 \lambda$. In these examples, DM orbits either hit a centrifugal barrier at $r>R$ or they intersect the neutron surface when ${dr \over d \tau} > 0$; there are no orbits that just barely touch the NS surface. 

The solid lines in \Fig{fig: veff examples} correspond to conserved angular momentum $L = \gamma m b_\text{max} v$. For large $\alf$, as can be seen in the figure, $b_\text{max}$ and the coordinate of the barrier's peak, $r_\text{b}$, are the solutions to 
\beq
V_\text{eff, +}'(r_\text{b})|_{L=\gamma m  b_\text{max} v}  = 0, \qquad 
V_\text{eff, +}(r_\text{b})|_{L=\gamma  m b_\text{max} v}  = \gamma m. \label{eq: bmax numerical}
\eeq
For very large $\alf$ there are two solutions, corresponding to the inner and outer barriers, respectively. The maximum impact parameter of DM that intersects the NS is the solution with smallest $b_\text{max}$.

Using $v^2 \ll {G M \over R} \ll 1$ and $\lambda \gg R$, we find that a very good analytic approximation for $b_\text{max}$ is given by the minimum of 
\beq
b_\text{max, inner}  \approx {R \over v} \sqrt{{{2 G M \over R} \left( 1 +  \tilde{\alpha} e^{-R/\lambda}\right) +\left({G M \over R}  \tilde{\alpha} e^{-R/\lambda}\right)^2 \over \left(1 - {2 G M \over R} \right)}}  \\
 \label{eq: b inner}
\eeq 
and 
\def\bmaxout{b_\text{max, outer}}
\beq
\bmaxout \approx \sqrt{ \lambda x \left({2 G M \over v^2} + \lambda x \right)} ~~ \text{with}~~x \approx \log \left({  \tilde{\alpha} \over {\lambda v^2 \over G M} + {1 \over \log  \tilde{\alpha}} } \right) \gtrsim 2 + \log 2,
 \label{eq: b outer}
\eeq
so 
\beq
b_\text{max} = \text{min} \left(b_\text{max, inner},  \bmaxout \right).
\label{eq: b}
\eeq

\begin{figure}[h]
\includegraphics{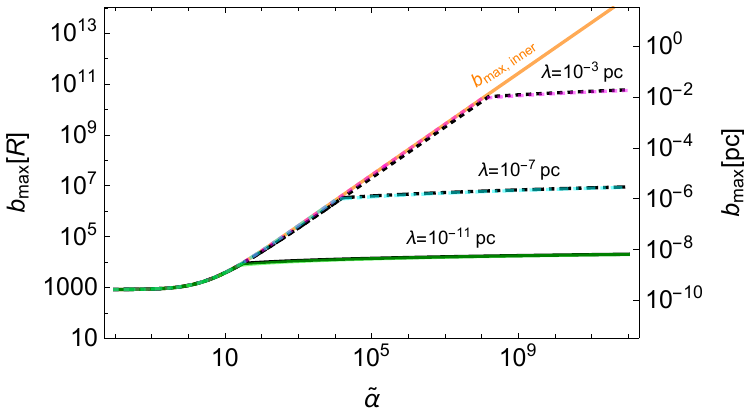}
\caption{Maximum impact parameter for which DM with asymptotic speed $v \sim 10^{-3}$ intersects a NS with circumferential radius $R$, assuming ${G M \over R}=0.2$, as a function of $\tilde{\alpha}$. The quoted $\lambda$ values and right-hand scale assume $R=10 \km$. The black lines show (more exact) numerical results while the colored lines show our analytic approximation in \Eq{eq: b}, which agrees with the numerical results to within better than a factor of 2 in the entire range. The kink in the $b_\text{max}$ curves correspond to the point where the outer centrifugal barrier at $r > \lambda$ occurs at smaller $b$ than the inner barrier closer to $r \sim R$. C.f.~\Fig{fig: veff examples}.}\label{fig: bmax}
\end{figure}

\Fig{fig: bmax} shows our analytic approximation in \Eq{eq: b} (color) alongside numerical solutions to \Eq{eq: bmax numerical} (black)\footnote{When ${G M \over R} = 0.2$ and $\lambda \gg R$, $r_\text{p} \lesssim R$ when  $\alf \lesssim 6$, and $b_\text{max}$ is given exactly by \Eq{eq: b inner}. The exact cutoff at small $\alf$ depends somewhat sensitively on ${G M \over R}$.} as a function of $\alf$ for three choices of force range, $\lambda$, assuming $v \sim 10^{-3}$, $R = 10 \km$, and ${G M \over R} = 0.2$. The inner barrier controls $b_\text{max}$ at small $\tilde{\alpha}$. For $\tilde{\alpha} \gg 1$, $b_\text{max}$ grows in proportion to $\tilde{\alpha}$ until it is cut to logarithmic growth because of the outer centrifugal barrier at radial distances greater than $\lambda$.

We end this appendix with a few comments on our analytic approximation and the behavior of $b_\text{max}$. \Eq{eq: b inner} is the impact parameter for which the effective potential equals $E$ at the NS surface in the $v \ll 1$ limit---i.e. $b(R)$ from \Eq{eq: b of rmin}.  It is only a slight overestimate of the impact parameter for which the inner centrifugal barrier peaks at $E$ (c.f. \Fig{fig: veff examples}), and it is exact in the limit $\alf \rightarrow 0$. 
In the opposite limit and when $\lambda \gg R$, $b_\text{max, inner} \approx {G M \over v}  \tilde{\alpha}$; the effective radius of the NS is approximately the geometric mean of the Coulomb classical and circular radii ($G M \tilde{\alpha}$ and ${G M \tilde{\alpha} \over v^2}$, respectively). \Eq{eq: b outer} comes from analyzing the effective potential in the ${G M \over r}, {(L/m)^2 \over r^2} \ll 1$ limit (the nonrelativistic limit), and $r_\text{p} \approx \lambda x $. For $x \lesssim 2$, the outer barrier does not exist. When the Yukawa force is much smaller than the gravitational force at the gravitational circular radius, $\tilde{\alpha} e^{-G M / v^2 \lambda} \ll 1$, gravity is important in determining $\bmaxout$, and $\bmaxout \rightarrow {1 \over v} \sqrt{2 \lambda G M \log (\tilde{\alpha} \log \tilde{\alpha})}$. In the opposite extreme, when $\tilde{\alpha} e^{-G M  / v^2  \lambda} > 1$, gravity is unimportant in determining the location of the barrier, and $\bmaxout \rightarrow \lambda \log\left({\tilde{\alpha} G M \over \lambda v^2}\right)$. This latter case is the only one where $b_\text{max} v $ evolves appreciably as a function of $v$; in all other limits discussed, $b_\text{max} v$ is essentially constant.\footnote{When $v \ll 1$, we expect $(L/m) \approx b v $ to be nearly independent of $v$ for orbits that sweep near the NS surface because the kinetic energy at closest approach is essentially independent of $v$; it is instead dominated by the fifth force and gravitational potential energy loss.} When computing averages over the asymptotic DM velocity distribution, we use the fact that $b v$ is nearly constant or---in the limit just discussed---mildly increasing (less than linearly) as a function of $v$.


\section{Timescale for heating once captured}\label{sec: timescale for heating}

In this appendix we estimate the period of the first obit after DM is captured by depositing energy $\Delta E_{\mathrm{graze}}$ during a transit of a NS. Our approach is inspired by that in \cite{2014}. As long as the apastron of this orbit, $r_\text{max}$, is large so that $\tilde{\alpha} {G M \over r_\text{max}} e^{-r_\text{max}/\lambda}, {G M \over r_\text{max}} \ll1$, a non-relativistic treatment  yields a good estimate because then the DM is non-relativistic on parts of its trajectory near the apastron, where it spends the majority of its time.  We proceed with a non-relativistic analysis.

The apastron is related to the DM's conserved non-relativistic energy, $E$, and orbital angular momentum magnitude, $L$, through
\beq
E = -{G M m_X \over r_{\text{max}}}(1 + \tilde \alpha e^{-r_{\text{max}}/\lambda}) + {1 \over 2}{L^2 \over m_X r_{\text{max}}^2}.
\eeq  And the period of an orbit with apastron $r_\text{max}$ and periastron $r_\text{min}$ is 
\beq
\Delta t = 2\int_{r_\text{min}}^{r_\text{max}} {dr \over \dot r} =  2 \int_{r_\text{min}}^{r_\text{max}} \left({2 E \over m_X}+ {2 G M \over r} \left( 1 + \tilde \alpha e^{-r/\lambda}\right) - {(L/m_X)^2 \over r^2} \right)^{-1/2} dr.
\eeq
Assuming orbital angular momentum doesn't increase upon capture,  $L_\text{closed orbit} \leq m_X b_\text{max} v$, and the orbital angular momentum term negligibly affects the period of orbits with $r_\text{max} \ggg R$, which are the orbits with the largest contribution to the total energy deposit timescale, we get
\beq
\Delta t \approx  2 \int_0^{r_\text{max}} \left({2 G M (1+ \tilde{\alpha} e^{-r/\lambda}) \over r} - {2 G M (1+\tilde{\alpha} e^{-r_\text{max}/\lambda}) \over r_\text{max}} \right)^{-1/2} dr.
\eeq
The above equation is the period of a maximally eccentric orbit. It reduces to Kepler's third law, $\Delta t = 2 \pi {a^{3/2} \over \sqrt{G M}}$ for a maximally eccentric orbit, where $a = r_\text{max} / 2$, when the ratio of the fifth force to the gravitational force at the apastron is very small, $\tilde \alpha (1 + {r_\text{max} \over \lambda})e^{-r_\text{max} / \lambda} \ll 1$. To extremely good accuracy when the force ratio at the apastron is either very small or very large, we find the integral can be approximated as
\beq
\Delta t \approx { \Delta t_\circ (r_\text{max}) \over 2 \sqrt{2}} =  {\pi \over \sqrt{2}} {r_\text{max}^{3/2} / \sqrt{G M} \over \sqrt{1 + \tilde{\alpha}\left(1 + {r_\text{max} \over \lambda}\right) e^{-r_\text{max}/\lambda}}} \label{eq: orbit period}
\eeq
where $\Delta t_\circ$ is the period of a circular orbit with radius $r_\text{max}$.  \Eq{eq: orbit period} is an overestimate. The overestimate is most significant when the force ratio at the apastron is comparable to 1 and $\tilde{\alpha}$ is large, but it comes within a factor of $4$ as long as $\tilde{\alpha} \lesssim 10^{30}$.

The apastron of the orbit after the first encounter, $r_{\text{max}, 1}$, is given by the solution to
\beq
{1 \over 2} m_X v^2 - \Delta E_{\mathrm{graze}} = -{G M m_X \over r_{\text{max}, 1}} \left( 1 + \tilde \alpha e^{-r_{\text{max}, 1}/\lambda} \right) + {m_X \over 2} {(L_1/m_X)^2 \over r_{\text{max}, 1}^2}. \label{eq: rmax1}
\eeq
A good estimate of the period of the first orbit after capture, $\Delta t_1$, is given by \Eq{eq: orbit period} evaluated at $r_{\text{max}}=r_{\text{max}, 1}$ determined by  \Eq{eq: rmax1} with $L_1 = 0$ (a maximally eccentric orbit). We note that $r_{\text{max}, 1}$ is an increasing function of $v$ and 
$r_{\text{max},1}(v) < r_{\text{max}, 1}(0) / (1 - v^2/v_\text{max}^2)$  so that $\Delta t (v) \lesssim {\Delta t(0) \over (1 - v^2/v_\text{cap}^2)^{3/2}}$.   For asymptotic speeds up to about $0.8\, v_\text{cap}$, $\Delta t_1(v) \approx \Delta t_1(0)$ is a good estimate; the period rapidly asymptotes to infinity thereafter, as $v \rightarrow v_\text{cap}$. 

Altogether, we have,
\beq
\Delta t_1(0) \approx GM {\pi \over \sqrt{2}} \left({m_X \over \Delta E_{\mathrm{graze}}} \right)^{3/2} {\left(1 + \tilde \alpha e^{-r_\text{max,1}/\lambda}\right)^{3/2} \over \sqrt{1 + \tilde \alpha \left(1 + {r_\text{max,1} \over \lambda}\right)e^{-r_\text{max,1}/\lambda}}} \label{eq: period}
\eeq where we redefined $r_\text{max,1}$ to be defined through \Eq{eq: rmax1} with $v=0, L_1=0$:
\beq
{\Delta E_\text{graze} }={G M m_X \over r_\text{max, 1}}(1+\alf e^{-r_\text{max, 1}/\lambda}).
\eeq 
When $r_\text{max,1} \gg \lambda$ and ${r_\text{max,1} \over \lambda} \tilde{\alpha} e^{-r_\text{max,1}/\lambda} \ll 1$, the period of the first orbit after capture is approximately $\Delta t_1 \approx G M {\pi \over \sqrt{2}} \left({m_X \over \Delta E_{\mathrm{graze}}} \right)^{3/2}$. When $r_\text{max, 1} \sim {G M m_X (1+\alf)\over \Delta E_\text{graze}} \ll \lambda$, $\Delta t_1 \approx G M {\pi \over \sqrt{2}} \left({m_X \over \Delta E_{\mathrm{graze}}} \right)^{3/2} (1 + \tilde \alpha)$.

The timescale is the limiting factor in the tidal heating rate when $\Delta t_1 \gtrsim t_\text{NS}$. From \Eq{eq: orbit period}, $r_\text{max,1} > \left({\sqrt{2 G M} \over \pi} \Delta t_1(0) \right)^{2/3} $ so $\Delta t_1 \gtrsim t_\text{NS}$ requires $r_\text{max} \gtrsim \left(t_\text{NS} \sqrt{G M} \right)^{2/3} \sim 10 \pc  \left({t_\text{NS} \over \Gyr}\right)^{2/3}$. For $\lambda \ll \left(t_\text{NS} \sqrt{G M} \right)^{2/3} $ and $\tilde{\alpha} < {\lambda \over \left(t_\text{NS} \sqrt{G M} \right)^{2/3} } e^{\left(t_\text{NS} \sqrt{G M} \right)^{2/3} /\lambda}$, we see $\Delta t_1 = t_\text{NS}$  corresponds to the contour $\Delta t_1 =  G M {\pi \over \sqrt{2}}\left({m_X \over \Delta E_\text{graze}}\right)^{3/2} = t_\text{NS}$.


\section{Pulsar Glitches} \label{app: glitches}

\def\Hz{\text{Hz}}
\def\glitchrate{\dot N_\text{glitch}}

In principle DM can transfer both energy and angular momentum to pulsars. At most, DM can transfer its entire angular momentum, $m_X b v$, during a close-range interaction. This would cause a typical  shift in the pulsar frequency---a glitch---of at most
\beq
 {\Delta \nu} \approx {|\Delta L_\text{NS}| \over I_\text{NS}} \approx {m_X \langle b_\text{max} v \rangle \over {2 \over 5} M R^2} \approx 10^5 \left( {m_X \over M} \right) \left({\langle b_\text{max} v \rangle \over R}\right) \left({10 \km \over R} \right) \Hz.
\eeq
 Meanwhile, the DM glitch rate per NS is at most the DM flux, given by, \beq
\glitchrate \lesssim {\rhoDM \over m_X} \pi \langle b_\text{max}^2 v \rangle  \sim   {10^{-24} \over  \yr}  {R^2 \over (10 \km)^2}{\rhoDM \over 0.4 \GeV/\cm^3}{1.4 M_\odot \over M}{10^{-3} \over v_p}\left({\langle b_\text{max}^2 v \rangle v_p \over R^2}\right){M \over m_X}. \label{eq: glitch rate}
\eeq 
 Consider $\Delta \nu \sim 10^{-9} \Hz$ as a benchmark.\footnote{Several convincing detections of frequency shifts as low as order $\Delta \nu \gtrsim 10^{-9} \Hz$ have been made~\cite{Abdo_2011, Basu_2021, Livingstone_2009}. See Ref.~\cite{Espinoza_2014} for a discussion of thresholds.}  Given gravity only, $\langle b_\text{max} v \rangle  / R \sim 1$, requiring $m_X / M \gtrsim 10^{-14}$ for a glitch of order $10^{-9} \Hz$ or greater, and the rate for such glitches is order $\glitchrate \sim {10^{-10} \over \yr}$ or less for the fiducial parameters in \Eq{eq: glitch rate}.  

However, the presence of a DM-NS fifth force with range greater than order a thousand kilometers opens up the possibility for larger $b_\text{max}$;  thus smaller $m_X$ could cause detectable glitches at greater rates than in the gravity-only case just discussed. Given maximal kinetic heating, the coldest observed NS constrains $\tilde{\alpha} \lesssim 10^2$ for $\lambda \gtrsim 10^{-10} \pc$ (see \Eq{eq: f=1 limit, large lam}), corresponding to ${\langle b_\text{max} v \rangle \over R} \lesssim {G M \over R} 10^2 \sim 20$. To get typical glitches of order $10^{-9} \Hz$ or greater, one would require ${m_X \over M} \gtrsim 10^{-15}$ and thus a glitch rate less than about $10^{-6}$ per NS per year. 

To conclude, in principle, a DM-baryon fifth force may have given rise to an interesting DM-induced pulsar glitch phenomenon, but kinetic heating limits rule this out.


\bibliography{DM-B-FifthForceConstraintsRefs.bib}

\end{document}